\documentclass[prd,byrevtex
,preprint
,nofootinbib,
,tightenlines
,showkeys
,showpacs]{revtex4}
\usepackage{amsmath,amsxtra,amssymb}
\usepackage[pdftex]{graphicx} 
\usepackage{float}
\newcommand{\bs}[1]{\boldsymbol{#1}}
\newcommand{\pa}{\partial}

\newcommand{\del}{\delta}

\begin{document}
\title{Viewing Black Holes by Waves}
\author{Ken-ichiro Kanai}
\email{kanai@gravity.phys.nagoya-u.ac.jp}
\author{Yasusada Nambu}
\email{nambu@gravity.phys.nagoya-u.ac.jp}
\affiliation{Department of Physics, Graduate School of Science, Nagoya 
University, Chikusa, Nagoya 464-8602, Japan}

\date{June 15, 2013, ver. 0.99} 

\begin{abstract}
    We study scattering of waves by black holes.  Solving a massless
    scalar field with a point source in the Schwarzschild spacetime,
    waves scattered by the black hole is obtained numerically. We then
    reconstruct images of the black hole from scattered wave data for
    specified scattering angles. For the forward and the backward
    directions, obtained wave optical images of black holes show rings
    that correspond to the black hole glories associated with existence of
    the unstable circular photon orbit in the Schwarzschild
    spacetime.
\end{abstract}
\maketitle

\section{Introduction}

Wave scattering and resulting diffraction effect are related to
familiar phenomena such as rainbows and glories. These phenomena are
caused by scattering of light by water droplet in atmosphere and
analysis based on the wave optics is required to understand its
formation and quantitative nature~\cite{NussenzveigHM:CUP:1992}. The
methodology and the formalism of the wave scattering problem are
applied to wide fields in physics for the purpose of obtaining the
information of the scatterer by analyzing the date of the scattering
wave.  For the gravitational physics, the wave related properties of
black hole spacetimes have been investigated by many researchers for
more than 40 years to obtain physical nature of black holes
\cite{ReggeT:PR108:1957,MatznerRA:PRD31:1985,FuttermanJAH:CUP:1988,AnninosP:PRD46:1992,
  AndersonN:PRD52:1995,AndersonN:X0011025:2001,GlampedakisK:CQG18:2001,DolanSR:CQG25:2008}.
T.~Regge and J.~A.~Wheeler~\cite{ReggeT:PR108:1957} investigated the
wave equation for gravitational perturbations in black hole spacetimes
and have shown that the Schwarzschild black hole is stable against the
perturbations. For the spinning Kerr black hole, it was shown that the
incident wave is amplified if the wave satisfies an appropriate
condition. This phenomena is called super radiance and peculiar to the
Kerr spacetime accompanying with the dragging
effect~\cite{FrolovYP:1998}. The quasi-normal oscillation of black
holes is also wave related property and this normal mode is
characterized by parameters of black holes~\cite{FrolovYP:1998}.  As a
straightforward application of the wave scattering theory, cross
sections of black holes are obtained from the scattering waves and the
nature of the black hole geometry is discussed from the view point of
waves.  Especially, peculiar to the wave scattering by the black hole,
the diffraction effect for the backward scattered wave becomes
significant and this leads to the phenomena so called the black hole
glories~\cite{MatznerRA:PRD31:1985,AnninosP:PRD46:1992}.  The main
purpose of the wave scattering problem by black holes is theoretical
understanding of the physics of black holes and wave propagation in
curved spacetimes. For astrophysical black holes, it is unlikely that
wave effects such as diffraction can be observed using current
technology, but there is a possibility that future technology enable
us to study interference effects in gravitationally lensed waves.

The treatment of the conventional wave scattering by black holes
mainly concerns analysis of the scattering amplitude and the cross
section.  In this paper, we consider wave scattering by black holes
from the view point of image formation. Our main motivation is to
investigate ``images'' of black holes formed by the incident wave
using wave optics~\cite{NambuY:JPCS410:2013,NambuY:IJAA3:2013}. In the
geometric optics, ``images'' of black holes are obtained by solving
null geodesics (light rays) in black hole spacetimes and we can draw
images of a black hole as ``black hole shadows''; if we assume a light
source behind a black hole is larger than the angular size of the
black hole, a distant observer see a dark spot that corresponds to the
apparent image of the black hole.  We expect to obtain the shadow
image of the black hole in the framework of wave scattering problems.
However, obtaining images using waves is not trivial task at first
sight; as is known, the scattering amplitude at the observer shows
diffraction pattern due to interference between scattered
waves. However, this diffraction pattern is not itself images of the
black hole. The theory of image formation in wave optics tells that we
must decode scattered waves using imaging device to reconstruct
images~\cite{SharmaKK:AP:2006}.  The scattered waves by a black hole
contains information of the black hole geometry and it is possible to
reconstruct images of the black hole.  For this purpose, we introduce
a convex lens to the configuration of the standard wave scattering by
black holes and aim to reconstruct images from scattering waves.

In this paper, we consider scattering of massless scalar waves by a
Schwarzschild black hole. The massless scalar waves are adopted as the
benchmark treatment for wave scattering problems by black holes and we
do not consider polarization degrees of freedom that is necessary for
the electro magnetic waves and gravitational waves.  We investigate
images of the black hole using wave optics.  As the source of incident
waves, we prepare a point source of the wave placed near the black
hole. This models the astrophysical black holes with active emission
regions such as accretion disks around them.  We solve the wave
equation for the massless scalar field (Helmholtz equation) using the
finite difference method.  Then, we reconstruct images from scattering
wave data and investigate characteristics of the black hole spacetime
appearing in images. This paper is organized as follows. In Sec.~II,
we introduce basic equations and our numerical method for solving the
scalar wave equation. In Sec.~III, we explain the method of images
reconstruction from scattering waves. In Sec.~IV, we present our
numerical results and Sec.~V is devoted to summary. We use units in
which $c=\hbar=G=1$ throughout this paper.

\section{Wave scattering in black hole spacetimes}
We solve the wave equation for the massless scalar field in the
Schwarzshild spacetime numerically. For this purpose, we rewrite the
Klein-Gordon equation for the scalar field to the Helmholtz equation
assuming that the wave field is monochromatic. Then by finite
differentiating the equation, we obtain the numerical solution of the
wave scattering problem.  The detail of the finite difference method
to solve the wave equation is presented in Appendix.

\subsection{Scalar wave equation in Schwarzschild spacetime}
For the  massless scalar field $\Phi$, the scalar wave equation in a curved
spacetime is 
\begin{equation}
\square\Phi=\frac{1}{\sqrt{-g}}\partial_{\mu}\left(\sqrt{-g}g^{\mu
      \nu}\partial_{\nu}\Phi\right)=S,
\label{eq:wave-eq}
\end{equation}
where $S$ is a source term of the wave. For the Schwarzschild
spacetime, the metric is
\begin{equation}
ds^2=-\left(1-\frac{2M}{r}\right)dt^2
+\left(1-\frac{2M}{r}\right)^{-1}dr^2+r^2(d\theta ^2+\sin^2\theta d\phi ^2),
\end{equation}
and the wave equation in the Schwarzschild spacetime is
\begin{align}
&-\left(1-\frac{2M}{r}\right)^{-1}\frac{\pa^2\Phi}{\pa t^2}
+\frac{1}{r^2}\frac{\pa}{\pa r}\left[r^2\left(1-\frac{2M}{r}\right)
    \frac{\pa\Phi}{\pa r}\right] \notag\\
&\qquad\qquad\qquad
+\frac{1}{r^2}\left[\frac{1}{\sin\theta}\frac{\pa}{\pa\theta}\left(\sin\theta
        \frac{\pa\Phi}{\pa\theta}\right)+\frac{1}{\sin^2\theta}
    \frac{\pa^2\Phi}{\pa\phi^2}\right]
=S(t,r,\theta,\phi). 
\label{eq:wave-eq1}
\end{align}
We consider the solution of this equation corresponding to the wave
scattering problem by the black hole; the wave is emitted by a point
source placed at $r=r_\text{S}, \theta=\pi$ (on the $\bar{z}$ axis)
and scattered by the black hole. The scattered wave is observed by a
detector placed at $r=r_\text{obs}$ (Fig.~1). We do not consider the
motion of the source in this paper. The location of the observer is
not restricted on the $\bar{z}$ axis. We assume that the wave is
monochromatic with the angular frequency $\omega$. This angular
frequency is defined with respect to the Schwarzschild coordinate time
$t$.
\begin{figure}[H]
\label{fig:bh-scatter}
\centering
  \includegraphics[width=0.4\linewidth,clip]{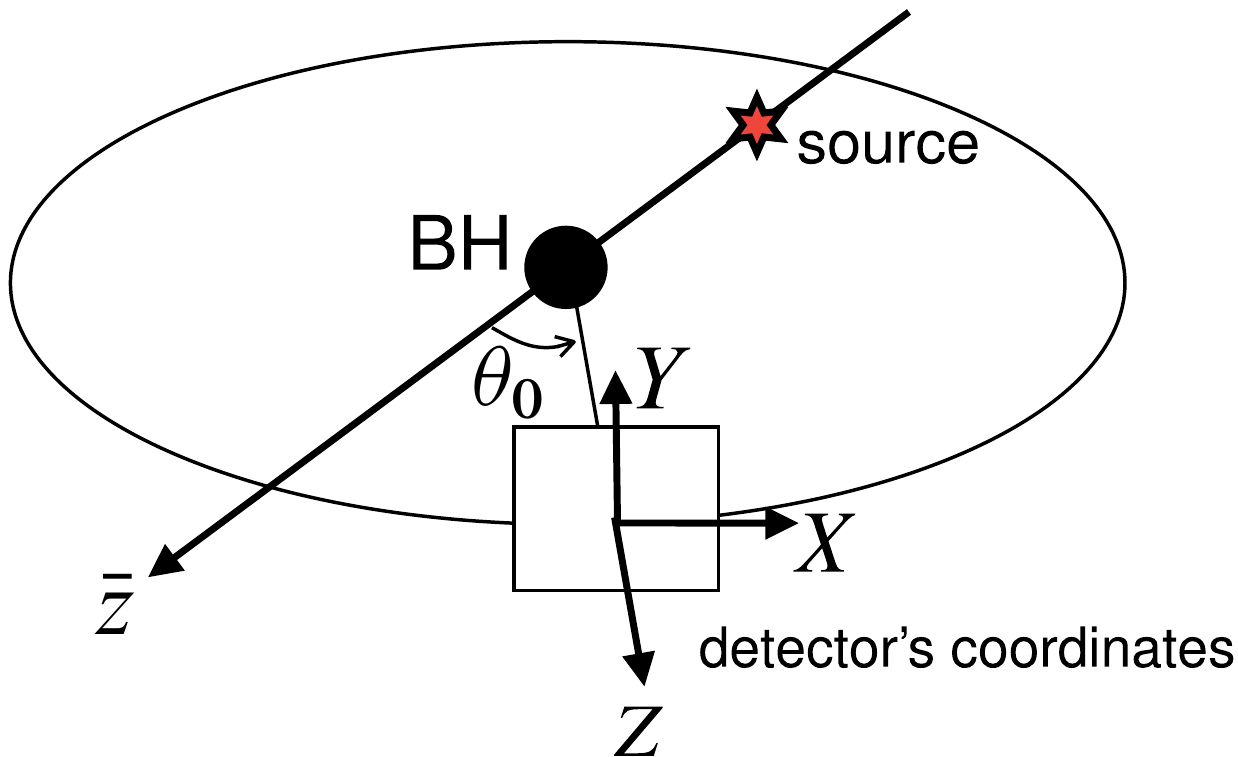}
  \caption{Configuration of the wave scattering problem by a black
    hole. We introduce spatial coordinates $(\bar{x},\bar{y},\bar{z})
    =(r\sin\theta\sin\phi,r\sin\theta\cos\phi,r\cos\theta)$. A point
    source is placed at
    $(\bar{x},\bar{y},\bar{z})=(0,0,-r_\text{S})$. The scattered waves
    are observed by a detector at $(\bar{x},\bar{y},\bar{z})
    =(r_\text{obs}\sin\theta_0, 0,
    r_\text{obs}\cos\theta_0)$, where $\theta_0$ corresponds to the
    scattering angle.}
\end{figure}
Because the Schwarzschild geometry is spherically symmetric, the
source and the black hole system of our scattering problem has an
axial symmetry.  Thus we can impose the axial symmetry for the wave
configuration: $\pa_\phi\Phi=0, \pa_\phi S=0$.  As we assume that the
time dependence of the scalar wave is $\Phi\propto e^{-i\omega t}$,
the wave equation can be written as the following Helmholtz equation:
\begin{equation}
    \frac{\pa^2\hat\Phi}{\pa x^{2}}+\frac{1}{r^2}
    \left(1-\frac{2M}{r}\right)\frac{1}{\sin\theta}\frac{\pa}{\pa\theta}
    \left(\sin\theta\frac{\pa\hat\Phi}{\pa\theta}\right)+\left[\omega^2
-\frac{2M}{r^3}\left(1-\frac{2M}{r}\right)\right]\hat\Phi
=S(r-r_\text{S},\theta-\pi)
\label{eq:rt-wavefunc}
\end{equation}
where $\hat\Phi=r\Phi$ and the tortoise coordinate is introduced by
$x=r+2M\ln\left(r/2M-1\right)$. $S(r-r_\text{S},\theta-\pi)$
represents the point source at $r=r_s, \theta=\pi$. In terms of the
delta functions,
\begin{equation}
  S(r-r_\text{S},\theta-\pi)=\frac{1}{r}\left(1-\frac{2M}{r}\right)
  \del(r-r_\text{S})\del(\cos\theta+1).
\end{equation}

\subsection{Boundary conditions}
We must impose boundary conditions to obtain solutions of our wave
equation.  We first consider the boundary conditions at $r=2M$ and
$r=\infty$.  By the definition of black holes, the 
wave  is purely ingoing at the horizon $r=2M$. Near the
horizon, the Helmholtz equation becomes
\begin{equation}
\pa_{x}^{2}\hat{\Phi}+\omega^2\hat\Phi=0.
\end{equation}
This equation contains the derivative with respect to the radial
coordinate only, and the wave propagates perpendicular to the
horizon. Taking into account the purely ingoing condition at the
horizon, the wave goes through the horizon perpendicular to it. By
restoring the time dependence, this boundary condition can be
expressed as
\begin{equation}
\left.\left(\pa_{x}\hat\Phi
-\partial_t\hat\Phi\right)\right|_{x\rightarrow-\infty}=0.
\label{eq:horizon-boundary}
\end{equation}
On the other hand, the wave is purely outgoing at the spatial infinity
because the source of wave is located at the finite distance from the
black hole and no wave is coming from outside of the numerical
box. Thus we can assume the wave behaves approximately as a spherical wave
at the outer boundary. This boundary condition can be expressed as
\begin{equation}
\left.\left(\pa_{x}\hat\Phi
+\pa_t\hat\Phi\right)\right|_{x=x_\text{outer}}=0.
\label{eq:out-boundary}
\end{equation}
Of course, this boundary condition is not exact one and this
approximated condition works well only for sufficiently large
numerical box compared to the size of the black hole and the location
of the point source. In our numerical calculation,
$r_\text{outer}=20.5M$ and $r_\text{S}\leq 6M$. Although
Eq.~\eqref{eq:out-boundary}  provide an approximated boundary
condition, we think it is sufficient for our purpose  in this paper.

Next, we consider the boundary condition associated with the symmetry
of the system.  According to the axial symmetry of the scalar field
configuration, $\hat\Phi$ should be an even function with respect to the
angle variable $\theta$: $\hat\Phi(\theta)=\hat\Phi(-\theta),
0\le\theta\le\pi$. Hence, the boundary condition for the wave on $\bar{z}$
axis is
\begin{equation}
\left.\frac{\partial\hat\Phi}{\partial\theta}\right|_{\theta=0,\pi}=0.
\label{eq:theta-boundary}
\end{equation}
In Eq.~(\ref{eq:rt-wavefunc}), the first order derivative term
$\cot\theta \partial_{\theta}\hat\Phi$ is contained and its
coefficient diverges at $\theta=0,\pi$.  Around $\theta=0$,
considering the boundary condition \eqref{eq:theta-boundary},
\begin{equation}
\cot\theta\approx\frac{1}{\theta}+O(\theta),\quad
\hat\Phi'(\theta)\approx\hat\Phi''(0)\theta+O(\theta^{2})
\end{equation}
where $'$ represents the derivative with respect to $\theta$. Hence,
\begin{equation}
\cot\theta\hat\Phi'(\theta)\approx
\hat\Phi''(0)+O(\theta)
\end{equation}
and this term is regular at $\theta=0$. In the similar manner, we can
see that the equation \eqref{eq:rt-wavefunc} is regular at
$\theta=\pi$. After all, at $\theta=0,\pi$, Eq.~(\ref{eq:rt-wavefunc})
reduces to
\begin{equation}
  \partial_{x}^2\hat\Phi
  +\frac{2}{r^2}\left(1-\frac{2M}{r}\right)
  \partial_{\theta}^2\hat\Phi+\left[\omega^2
      -\frac{2M}{r^3}\left(1-\frac{2M}{r}\right)\right]\hat\Phi
  =S(r-r_\text{S},\theta-\pi).
\label{eq:rt-wavefunc2}
\end{equation}
We adopt this equation as the equation to determine the wave field on the
$\bar{z}$ axis ($\theta=0,\pi$).

Finally, we consider the boundary condition around the point source.
Let us consider a location P near the point source S.  In the
neighborhood of the point source, the asymptotic form of the wave
function in the lowest order of the WKB approximation is given by
  \begin{equation}
    \label{eq:waveS}
        \Phi_\text{P}=\frac{\hat\Phi_\text{P}}{r_\text{P}}\sim
        \frac{A}{\ell_\text{PS}}\exp\left(\frac{i\omega\ell_\text{PS}}{
            \sqrt{1-2M/r_\text{P}}}\right)
  \end{equation}
  where $A$ is a constant representing the strength of the source and
  $\ell_\text{PS}$ is the spatial proper distance between P and S:
\begin{equation}
  \ell_\text{PS}^2\approx
  \left(1-\frac{2M}{r_\text{P}}\right)^{-1}(r_\text{P}-r_\text{S})^2
  +r_\text{P}^2(\theta_\text{P}-\theta_\text{S})^2
  \approx\left(1-\frac{2M}{r_\text{P}}\right)(x_\text{P}-x_\text{S})^2
  +r_\text{P}^2(\theta_\text{P}-\theta_\text{S})^2.
\end{equation}
We use this asymptotic behavior of the wave to detemine the wave field
around the point source. 
\section{image formation in wave optics}

Most of previous works on wave scattering by black holes aim to obtain
the scattering amplitude or the differential cross section to
investigate wave effects. On the other hand, we want to consider images
constructed from the scattered waves by black holes.  By using images,
it is also possible to discuss the semi-classical nature of the wave
scattering by black holes.  We have introduced the method to
reconstruct images from waves in the gravitational lensing
system~\cite{NambuY:JPCS410:2013,NambuY:IJAA3:2013} and obtained images of
gravitational lensing in wave optics. We apply the same method to the
wave scattering by black holes. In the wave optics, the image
formation can be understood as diffraction effect of waves. For this
purpose, image formation devices such as a convex lens are introduced.
\begin{figure}[H]
    \centering
    \includegraphics[width=0.35\linewidth,clip]{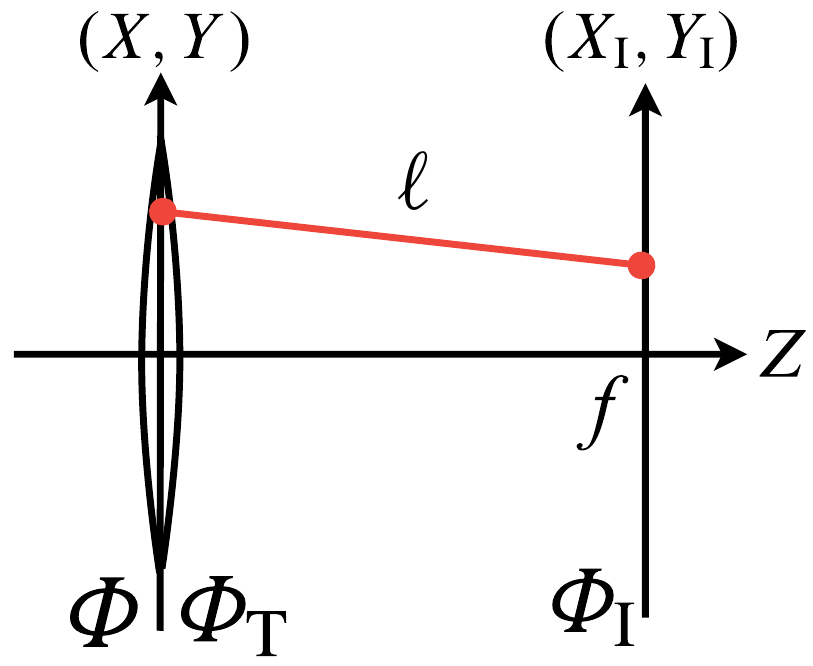}
    \caption{An image formation system with a convex lens. Incident
      waves are transformed by the lens and an image of a source
      object appears on the image plane. This image corresponds to the
      Fourier transformation of the incident wave.}
\end{figure}
In our analysis, we adopt a thin convex lens.  In this paper, we use a
word ``lens'' to represent a device taking the Fourier transformation
of the wave field and ``lens'' does not necessarily suggest a physical
artefact used in optics. Antennas of a VLBI telescope is also a kind of
``lens''. Let us $\Phi(\vec{X})$ be the incident wave in front of the
lens and $\Phi_{\text{T}}(\vec{X})$ is the transmitted wave by the
lens. $\vec{X}=(X,Y)$ denotes the location of a point on the lens
plane $Z=0$. These two wave functions are connected by the following
relation
\begin{equation}
  \Phi_{\text{T}}(\vec{X})=e^{-i\omega\frac{|\vec{X}|^2}{2f}}\Phi(\vec{X}),
\end{equation}
where  $f$ represents the focal length of the lens. For a
point source placed at $Z=-f$ (front focal plane), the wave in front
of the lens is
$$
 \Phi(\vec{X})=e^{i\omega\sqrt{|\vec{X}|^2+f^2}}\approx
 e^{i\omega(f+\frac{|\vec{X}|^2}{2f})},
$$
where we assumed $|\vec{X}|\ll f$. Hence the transmitted wave becomes
$$
 \Phi_{\text{T}}(\vec{X})=e^{i\omega f}
$$
and its phase becomes independent of $(X,Y)$. This means that the
transmitted wave is the plane wave. Thus the convex lens converts a
spherical wave front to a plane wave front.  Now we consider the wave
amplitude on the focal plane $Z=f$ (image plane). Using the
Fresnel-Kirchoff diffraction formula \cite{SharmaKK:AP:2006}, the
wave on the image plane is given by the following diffraction
integral:
\begin{equation}
 \Phi_{\text{I}}(\vec{X}_\text{I})\propto\int_{|\vec{X}|\leq d}
    d^2\vec{X}\,\Phi(\vec{X})\,e^{-i\omega\frac{|\vec{X}|^2}{2f}}
    \times\frac{e^{i\omega\ell}}{\ell}   
\end{equation}
where $\ell$ is  path length between a point on the lens plane and
a point on the image plane (see Fig.~2) and $d$ is a radius of the
lens. Using $\ell=\sqrt{|\vec{X}-\vec{X}_{\text{I}}|^2+f^2}\approx
f+|\vec{X}-\vec{X}_{\text{I}}|^2/2f$,
\begin{equation}
    \Phi_{\text{I}}(\vec{X}_\text{I})
    \propto\int_{|\vec{X}|\leq d} d\vec{X}\,\Phi(\vec{X})\,
    e^{-\frac{i\omega}{f}(\vec{X}_\text{I}\cdot\vec{X})}. \label{eq:waveImage}
\end{equation}
Hence $\Phi_{\text{I}}$ is the Fourier
transformation of the incident wave $\Phi$.

Let us consider the situation that the incident wave can be written as
the WKB form
\begin{equation}
    \Phi(\vec{X})=A e^{i\omega S(\vec{X})}
    =A\exp\left[i\omega\left(S(\vec{X}_*)+\frac{1}{2}S''(\vec{X}_*)
         (\vec{X}-\vec{X}_*)^2+\cdots\right)\right]
\end{equation}
where $\vec{X}_*=(X_*, Y_*)$ denotes the location of the point that
the null geodesics intersects with $Z=0$ plane. This point corresponds
to the classical path of the null ray  obtained as the saddle point of
the action $S$, $S'(\vec{X}_*)=0$. The classical path is a null
geodesics connecting a source and a point on the lens plane. For this
form of waves, the wave amplitude on the image plane
\eqref{eq:waveImage} becomes
\begin{align}
    \Phi_\text{I}(\vec{X}_\text{I})&\propto
    \int_{|\vec{X}|\leq d}d^2\vec{X}
    \,\exp\left[i\omega\left\{\frac{S''}{2}(\vec{X}-\vec{X}_*)^2
-\frac{\vec{X_{\text{I}}}\cdot\vec{X}}{f}\right\}\right] \notag\\
 &\propto \frac{J_1\left(\omega
       d\left|S''\,\vec{X}_*+\vec{X}_\text{I}/f\right|\right)}{
   \omega d\left|S''\,\vec{X}_*+\vec{X}_\text{I}/f\right|}.
\end{align}
We have assumed $|\omega S''d^2|\ll 2\pi$ and omitted the quadratic
term $|\vec X|^2$ in the exponent to evaluate the integral at the
last step of the calculation; this is the condition for the Fraunhofer
diffraction. On the dimensional analysis, $S''\sim 1/r$ where $r$ is
the distance between the black hole and the observer. As $d\ll r$,
this condition is easily satisfied.  For $\omega d\gg 1$ (geometric
optics limit), we obtain
\begin{equation}
    \Phi_\text{I}(\vec{X}_\text{I})
\propto\del^2\left(\vec{X}_\text{I}+fS''\,\vec{X}_*\right).   
\end{equation}
Thus we recover a point image of the point source on the image plane;
the location of the image corresponds to the location determined by
the null geodesics. If we do not take geometric optics limit, the
image acquires diffraction effects and the image of the point source
has a fine size called the Airy disk.  We apply this model of image
formation to scattering problems by black holes and obtain images of
black holes (Fig.~1).  For a point $(X,Y)$ on the lens plane $Z=0$,
the angle between this point and the $\bar{z}$ axis is given by
\begin{equation}
\cos\theta=\frac{X}{r}\sin\theta_0+\sqrt{1-\left(\frac{X}{r}\right)^2
-\left(\frac{Y}{r}\right)^2}\,\cos\theta_0.
\end{equation}
We use this relation to evaluate the Fourier integral
\eqref{eq:waveImage} to obtain images.

As an example of image reconstruction from scattering waves, we
present the gravitational lensing by a point source. Assuming that the
gravitational field is weak, the wave equation for the massless scalar
field reduces to the Newtonian form~\cite{FuttermanJAH:CUP:1988}
\begin{equation}
  \nabla^2\Phi_\text{N}+\left(\omega^2
    +\frac{4M\omega^2}{r}\right)\Phi_\text{N}=0,  
\end{equation}
where $\nabla^2$ is the flat space Laplacian and the problem is
equivalent to that of the Coulomb scattering in quantum mechanics. The
exact wave function for the scattering problem with a plane wave incident
from infinity is given by
\begin{equation}
  \Phi_\text{N}(\theta_0)=e^{\pi\omega M}\Gamma(1-2i\omega M)e^{i\omega
    r\cos\theta_0}{}_1F_1(2i\omega M, 1; i\omega r(1-\cos\theta_0)).  
\end{equation}
The wave amplitude at $r=20M$ is shown in Fig.~3. For small scattering
angle, oscillatory behavior due to diffraction between the incident
wave and the scattered wave can be observed. The behavior of the
scattering amplitude for small scattering angle (forward direction) can
be obtained by the asymptotic behavior of the confluent geometric
function:
\begin{equation}
    \Phi_\text{N}(\theta_0)\propto J_0(2\omega\sqrt{Mr}\,\theta_0)\quad
    \text{for}\quad |\theta_0|\ll 1.
\end{equation}
The Fourier transformation \eqref{eq:waveImage} for this function
results in the function with a peak at the angular radius
$\theta_\text{E}\equiv\sqrt{4M/r}$, that corresponds to the Einstein
ring.  On the other hand, for large value of $\theta_0$, the
diffraction effect becomes small and a ring image does not appear in
the backward direction.
\begin{figure}[H]
  \centering
  \includegraphics[width=0.5\linewidth,clip]{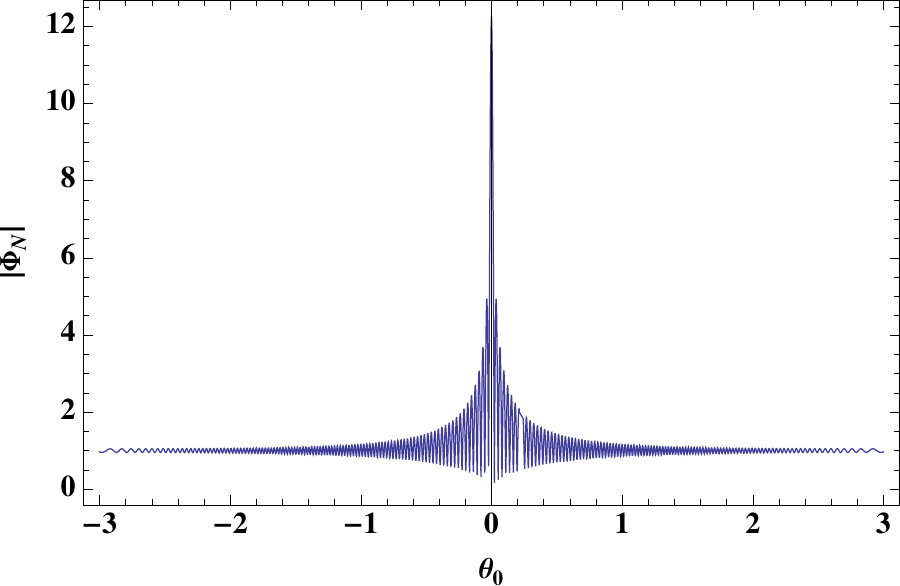}
  \caption{The wave amplitude $|\Phi_\text{N}|$ for $M\omega=12$ at $r=20M$.}
\end{figure}
The reconstructed images from the scattering wave using
\eqref{eq:waveImage} are shown in Fig.~4. For $\theta_0=0$, we obtain
an image of the Einstein ring with angular radius $\theta_E$. For
$\theta_0\neq 0$, double images of the point source appear. These
results reproduce images of the gravitational lensing obtained by the
ray tracing method using the lens equation.
\begin{figure}[H]
  \centering
  \includegraphics[width=0.35\linewidth,clip]{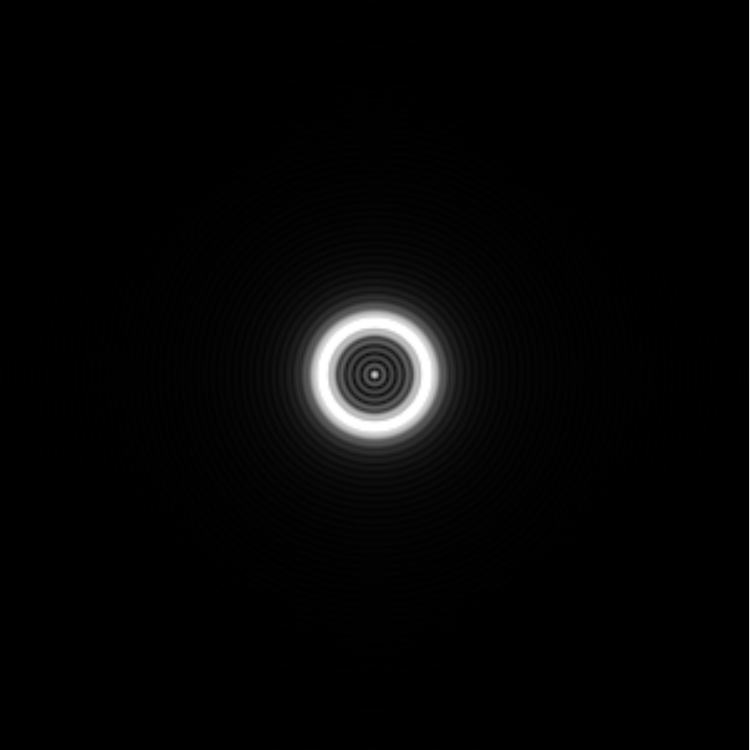}%
  \hspace{0.5cm}
  \includegraphics[width=0.35\linewidth,clip]{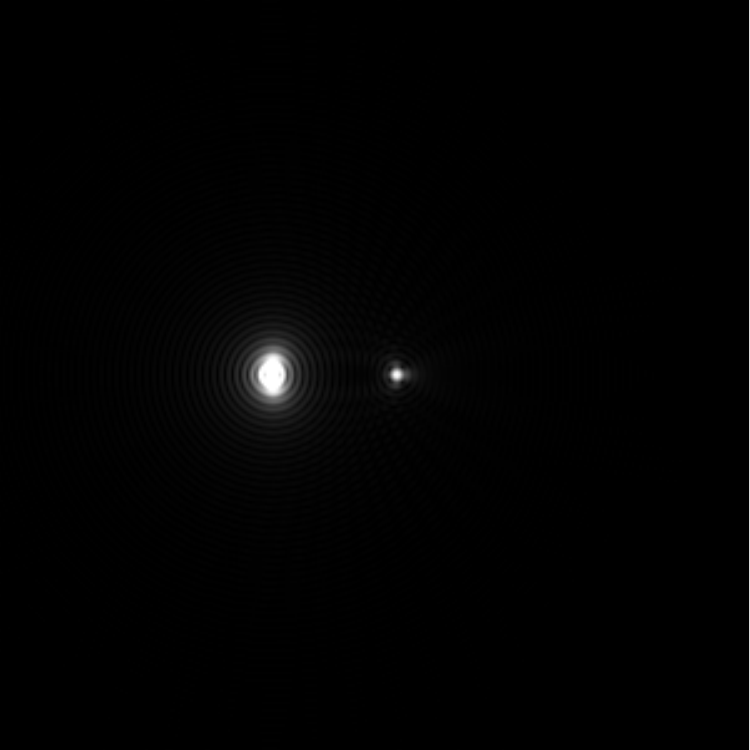}
  \caption{The left panel is the image for $\theta_0=0$ and this is
    the Einstein ring. The angular radius of the ring is given by
    $\theta_\text{E}$. The right panel is the image for
    $\theta_0=\pi/4$ and double images of the source appear. These
    images are obtained for $\omega M=12, r=20M, d/r=0.2$.}
\end{figure}

\newpage
\section{Numerical Results of Wave Scattering by Black Hole}

We numerically solved the massless field in the Schwarzschild
spacetime for the angular frequencies $M\omega=2,12,24$ and the source
positions $r_\text{S}=2.5M, 3M, 6M,15M$. The detector is located at
$r_\text{obs}=20M$.  The numerical grid size is $1001\times 1001$.  We
have also done the calculation with the grid size $501\times 501$ to
check the validity of our numerical results. In this paper, we only
present the result for $M\omega=2, 12$ and $r_\text{S}=2.5M, 6M$.

\subsection{$M\omega=12$ case}
We first present the result for $M\omega=12$ case.  The source
position is $r_\text{S}=6M$ and moderately far from the black
hole. This setting corresponds to the standard analysis of wave
scattering by black holes that the source is placed at spatial
infinity and the incident wave is treated as a plane wave.  At the
location of the observer, we do not distinguish the incident waves and
the scattered waves. Hence the detector receives both the incident
waves and the scattering waves, and their superposed wave amplitude is
obtained.  As the result, despite of the long range Coulomb like
nature of the gravitational field, ``scattering amplitude''
$|\Phi_\text{obs}|$ remains finite even for small scattering angles
(see Fig.~6). The situation is the same as the gravitational lensing
presented in the last section. The standard handling of the scattering
problem extracts purely scattered wave and results in diverging
scattering amplitude for small scattering angle (forward direction)
due to the long range nature of the gravitational force.

\begin{figure}[H]
  \centering
  \includegraphics[width=0.4\linewidth,clip]{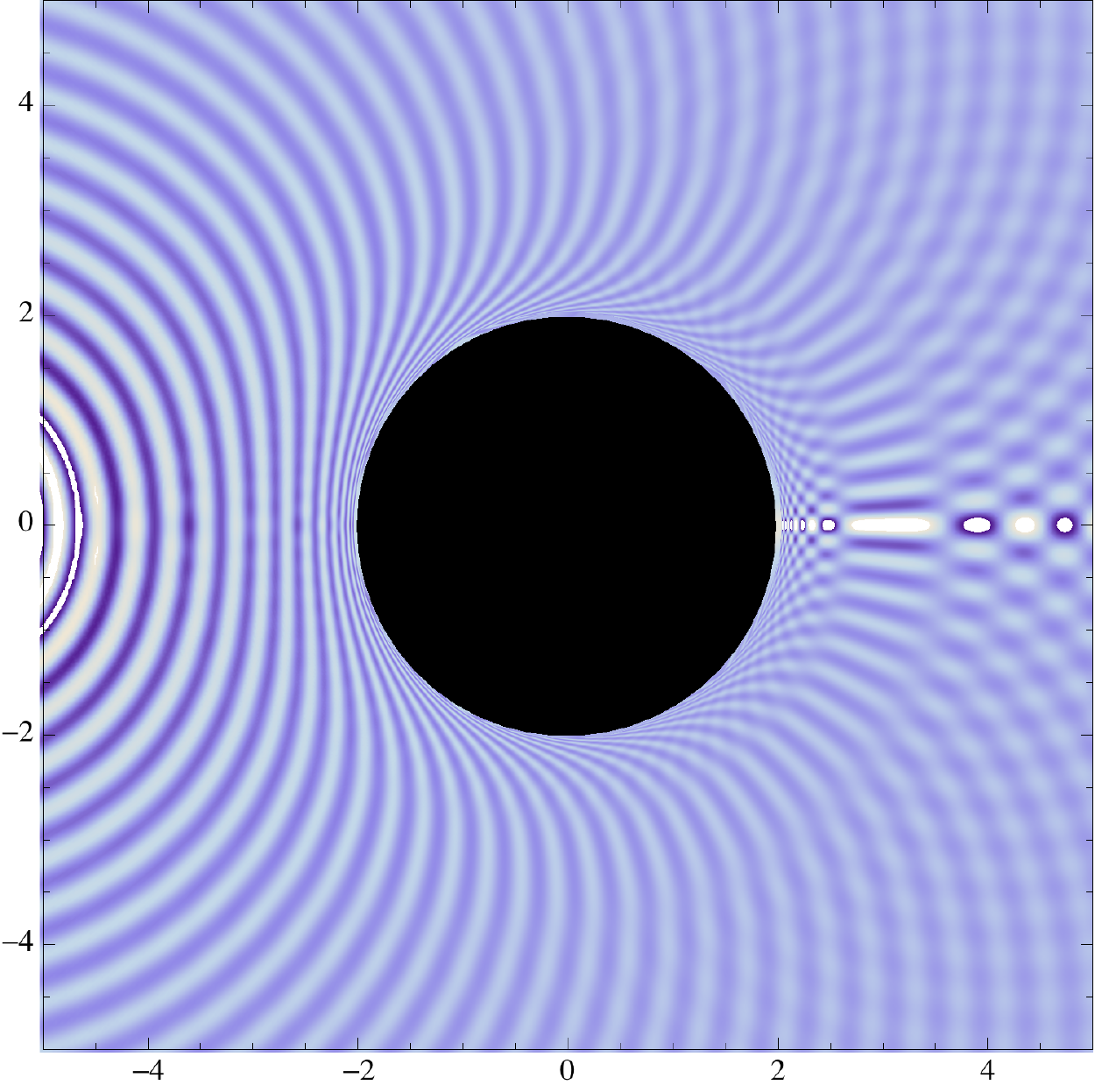}%
  \hspace{0.5cm}
  \includegraphics[width=0.4\linewidth,clip]{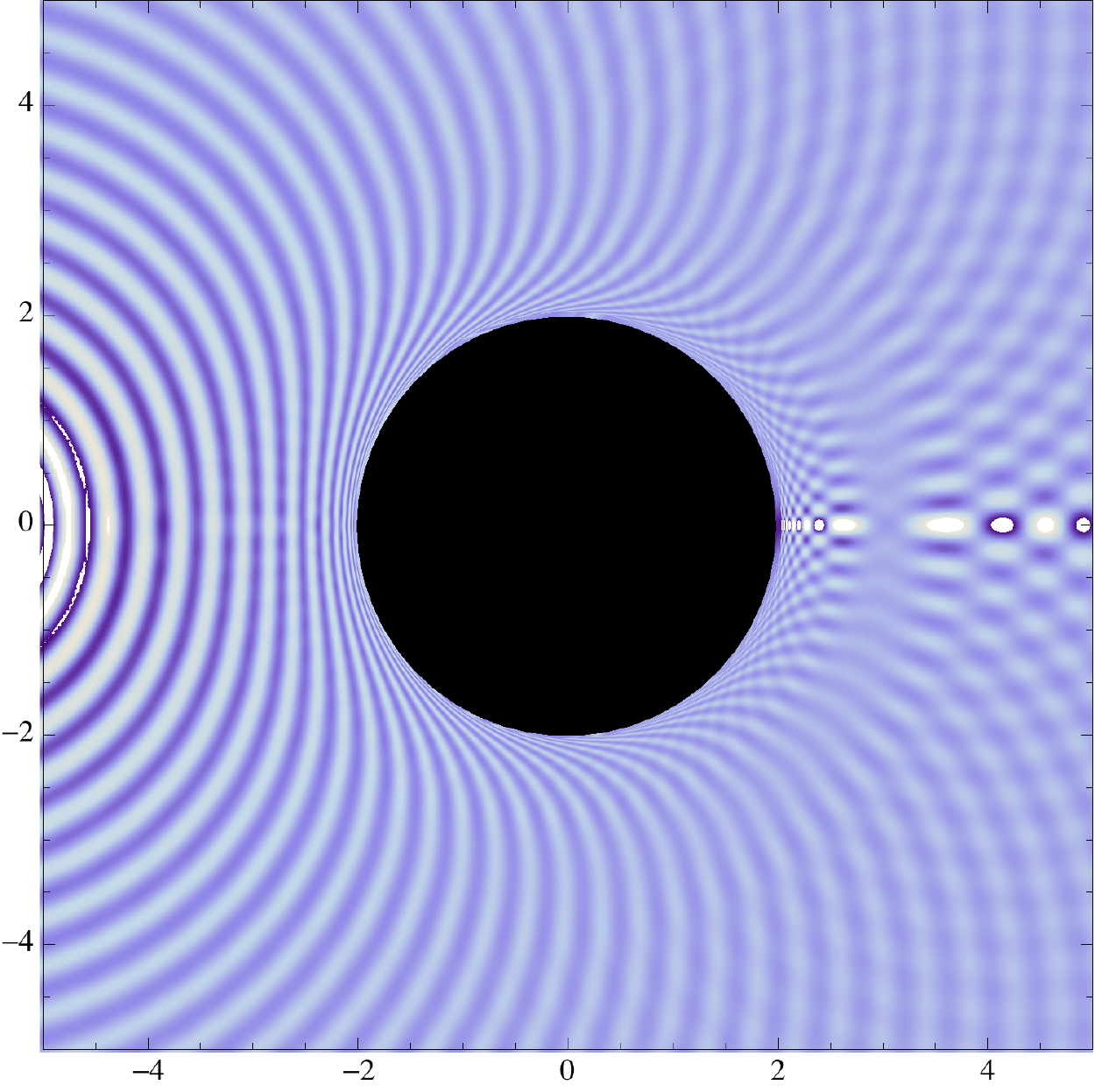}
  \caption{Spatial distribution of $\Phi$ for $M\omega=12$ on
    $(\bar{z},\bar{x})$
    plane. Left panel: $\text{Re}[\Phi]$. Right panel:
    $\text{Im}[\Phi]$. The point source is placed at
    $(\bar{z},\bar{x})
=(-6M,0)$.}
\end{figure}

Fig.~5 shows the distribution of $\Phi$ on $(\bar{z},\bar{x})$ plane. We can
observe that the wavelength of the incident wave becomes shorter near
the black hole horizon due to gravitational blue shift. Along the $\bar{z}$
axis ($\bar{z}>0$), the amplitude of the wave is enhanced due to constructive
interference of scattered waves. It is also possible to observe the
characteristic wave pattern  formed around the unstable orbit
$r=3M$.

\begin{figure}[H]
  \centering
  \includegraphics[width=0.5\linewidth,clip]{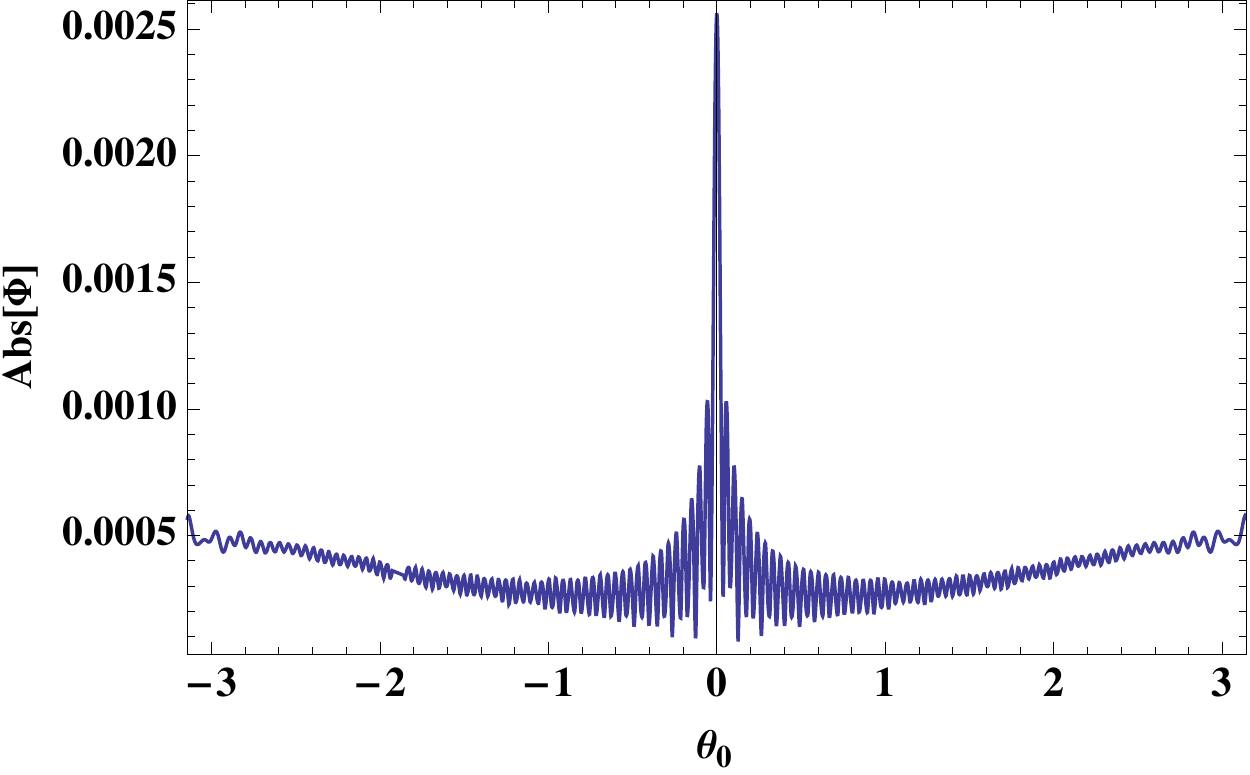}
  \caption{The observed wave amplitude at
     $r_\text{obs}=20M$. $M\omega=12, r_\text{S}=6M$.}
\end{figure}

As already mentioned, the wave amplitude remains finite for
$\theta_0\sim 0$. The wave amplitude has oscillatory behavior for all
values of the scattering angle. This is contrasted with the scattering
amplitude for the gravitational lensing. Except for $\theta_0\sim 0 $,
the wave amplitude increases as the scattering angle increases. As the
distance from the point source to the observer varies depending on the
scattering angle, the observed wave amplitude increases for the
backward direction because the distance between the source and the
observer decreases.  The oscillatory behavior of the wave amplitude at
the forward and the  backward directions corresponds to so called glory
effect~\cite{MatznerRA:PRD31:1985,AnninosP:PRD46:1992} and it is
possible to identify the glory in images in our analysis. For black
holes, glories arise because a wave can be deflected through an angle
greater than $\pi$ and the glory scattering is associated with the
unstable photon orbit at $r=3M$.

The reconstructed images from the scattering wave are shown in
Fig.~7. We have applied the formula \eqref{eq:waveImage} with
$d/r_\text{obs}=0.5$ to obtain these images. For $\theta_0=0$, the
image is a ring that corresponds to the forward glory. As is known,
the scattering amplitude for the forward and the backward glory is
given by~\cite{MatznerRA:PRD31:1985,AnninosP:PRD46:1992}
\begin{equation}
    \label{eq:Jg}
    \Phi(\theta_0)\sim J_0(\ell_g\theta_0)
\end{equation}
where $\ell_g=3\sqrt{3}M\omega$ corresponds to the critical impact
parameter that the incident null ray can escape to infinity. In terms
of images obtained by the Fourier transformation of
\eqref{eq:Jg}, the apparent angular radius of the ring is given by
\begin{equation}
    \label{eq:tg}
    \theta_g=\frac{3\sqrt{3}M}{r}.
\end{equation}
The solution \eqref{eq:Jg} and the relation \eqref{eq:tg} can be used
to check the numerical result.  For the backward direction
$\theta_0=\pi$, we cannot identify the ring image well because it is
drown in the wave of the point source. But in the plot of the
intensity distribution (Fig.~8), we can identify small dips of the
intensity at the radius corresponding to the ring of the backward
glory.

\begin{figure}[H]
  \centering
  \includegraphics[width=0.3\linewidth,clip]{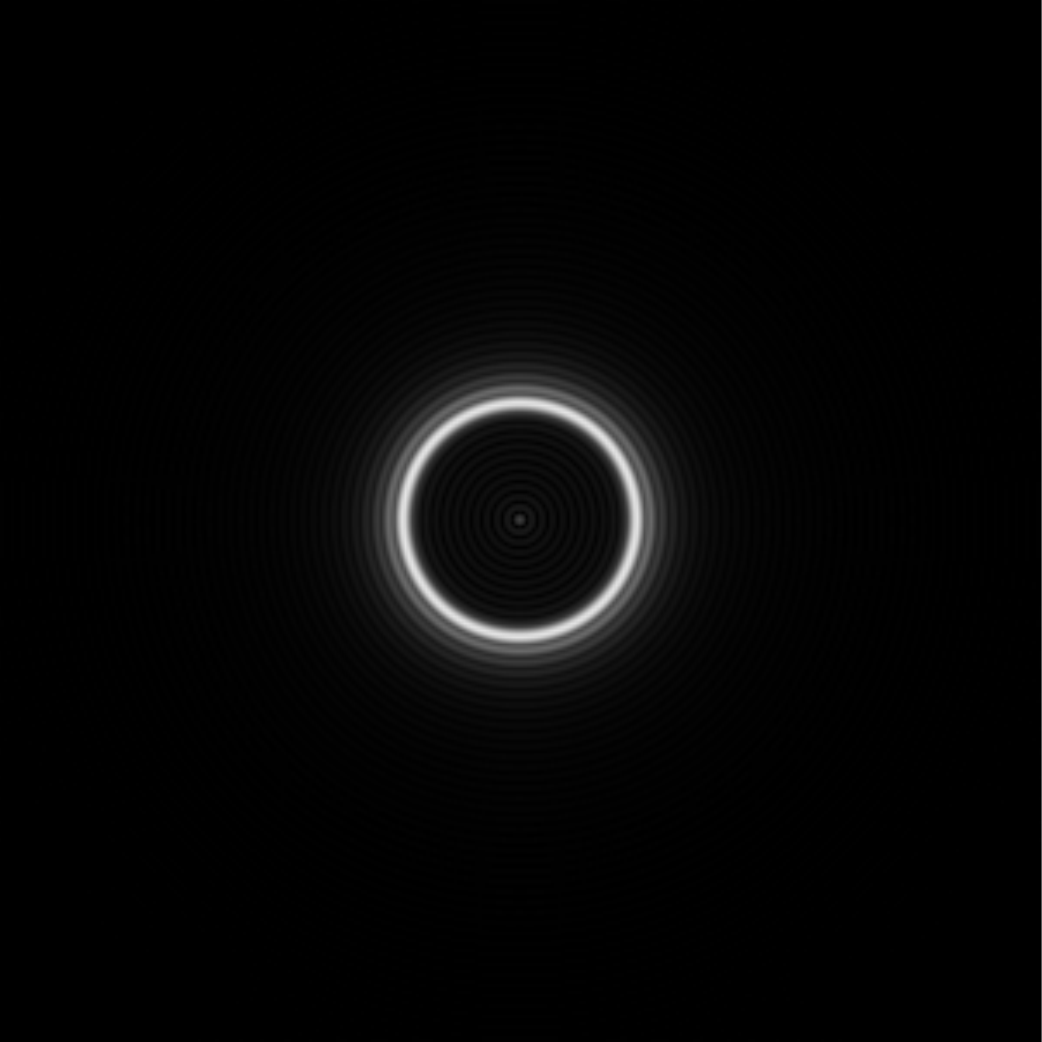}
  \includegraphics[width=0.3\linewidth,clip]{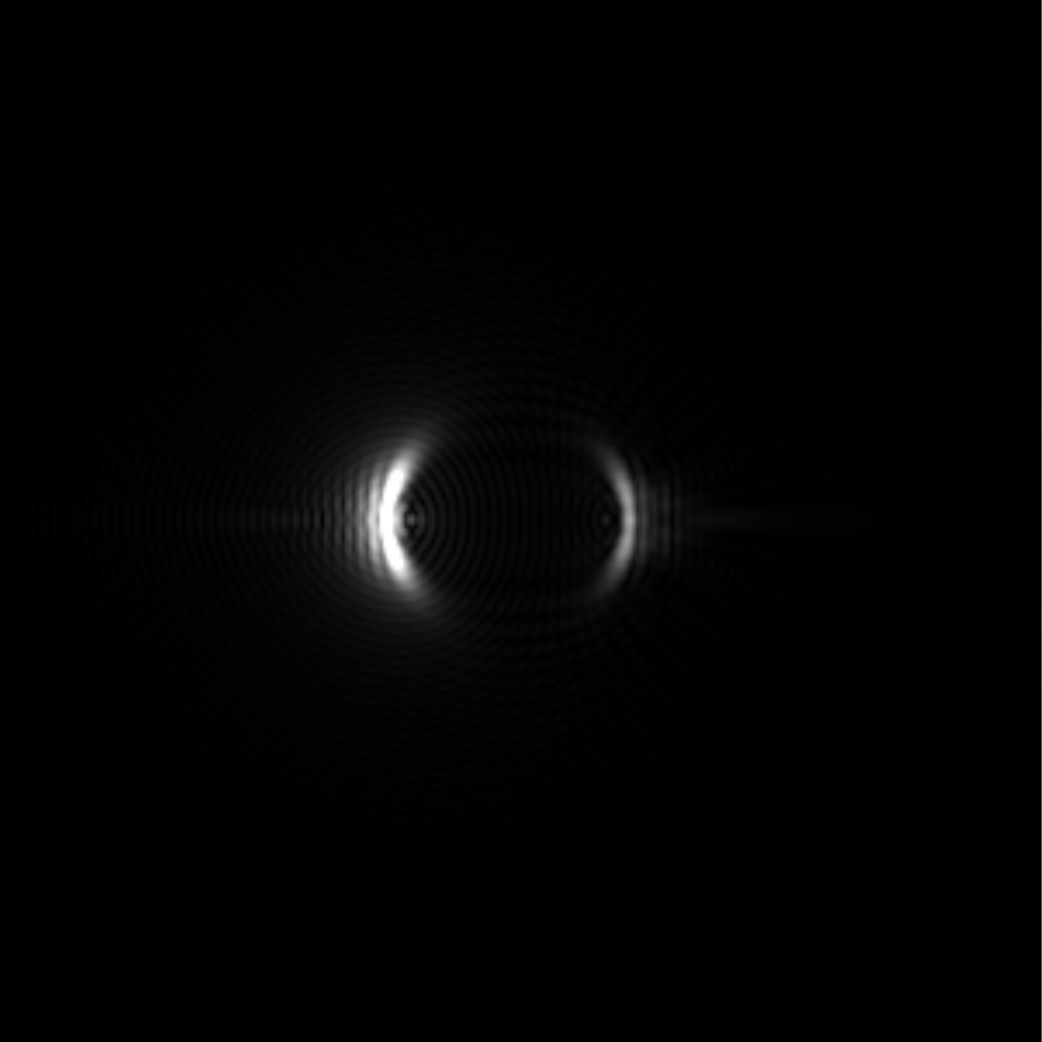}
  \includegraphics[width=0.3\linewidth,clip]{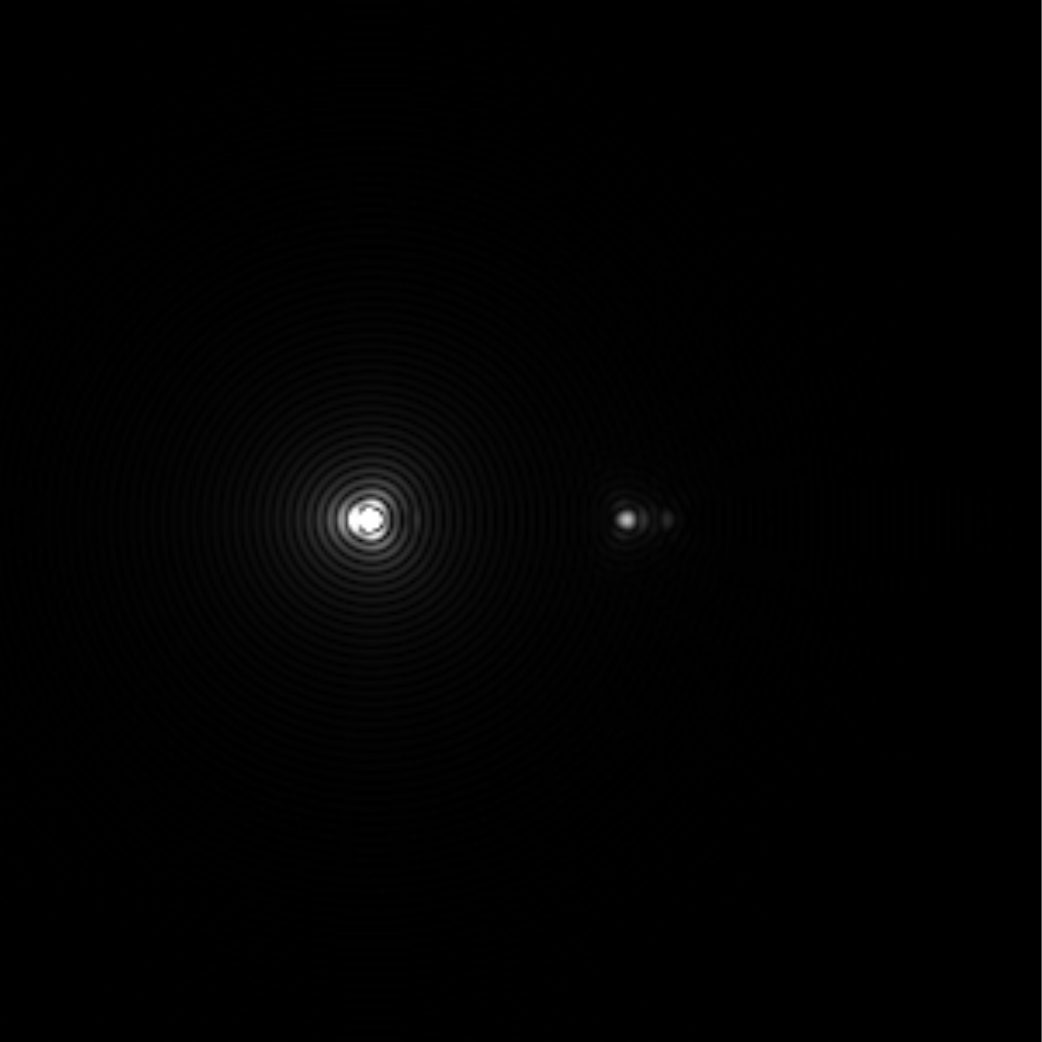}
  \includegraphics[width=0.3\linewidth,clip]{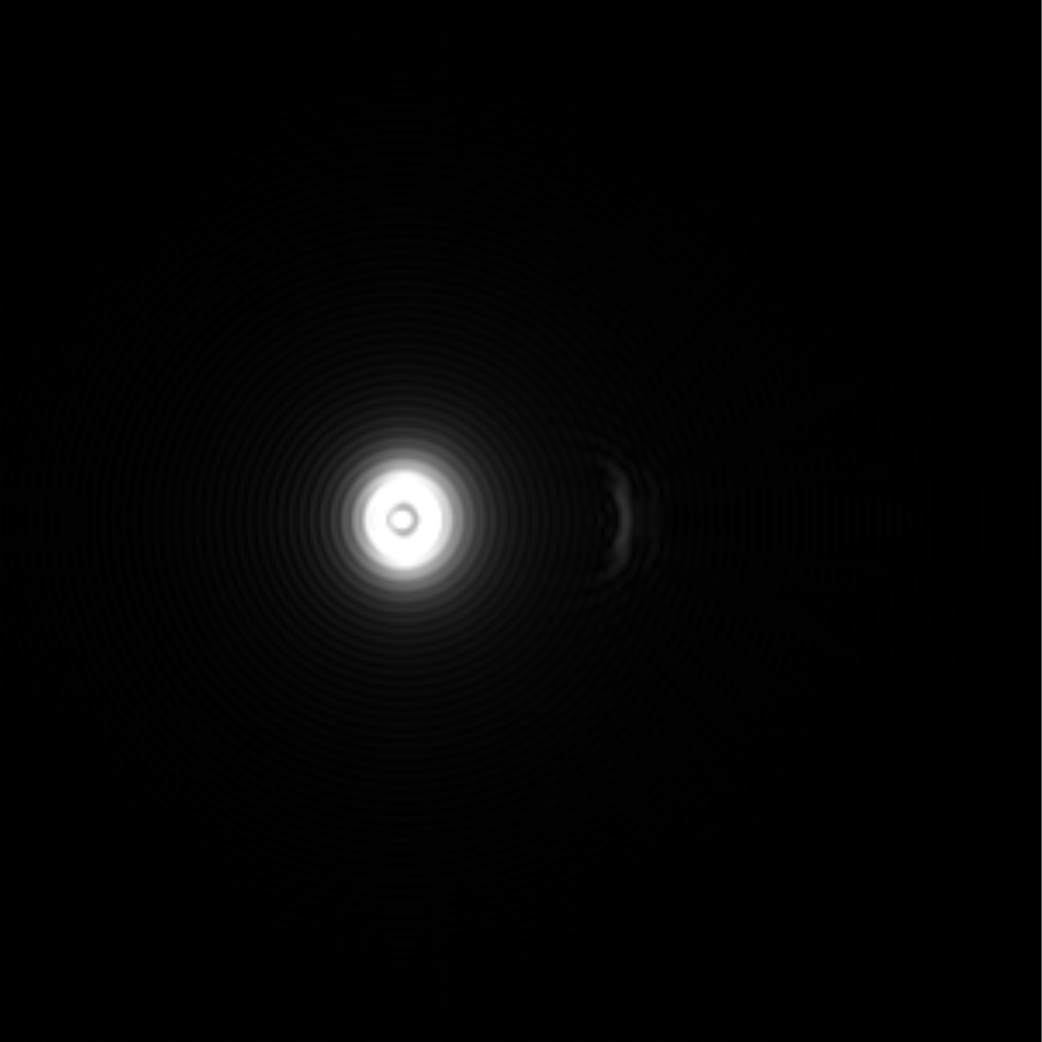}
  \includegraphics[width=0.3\linewidth,clip]{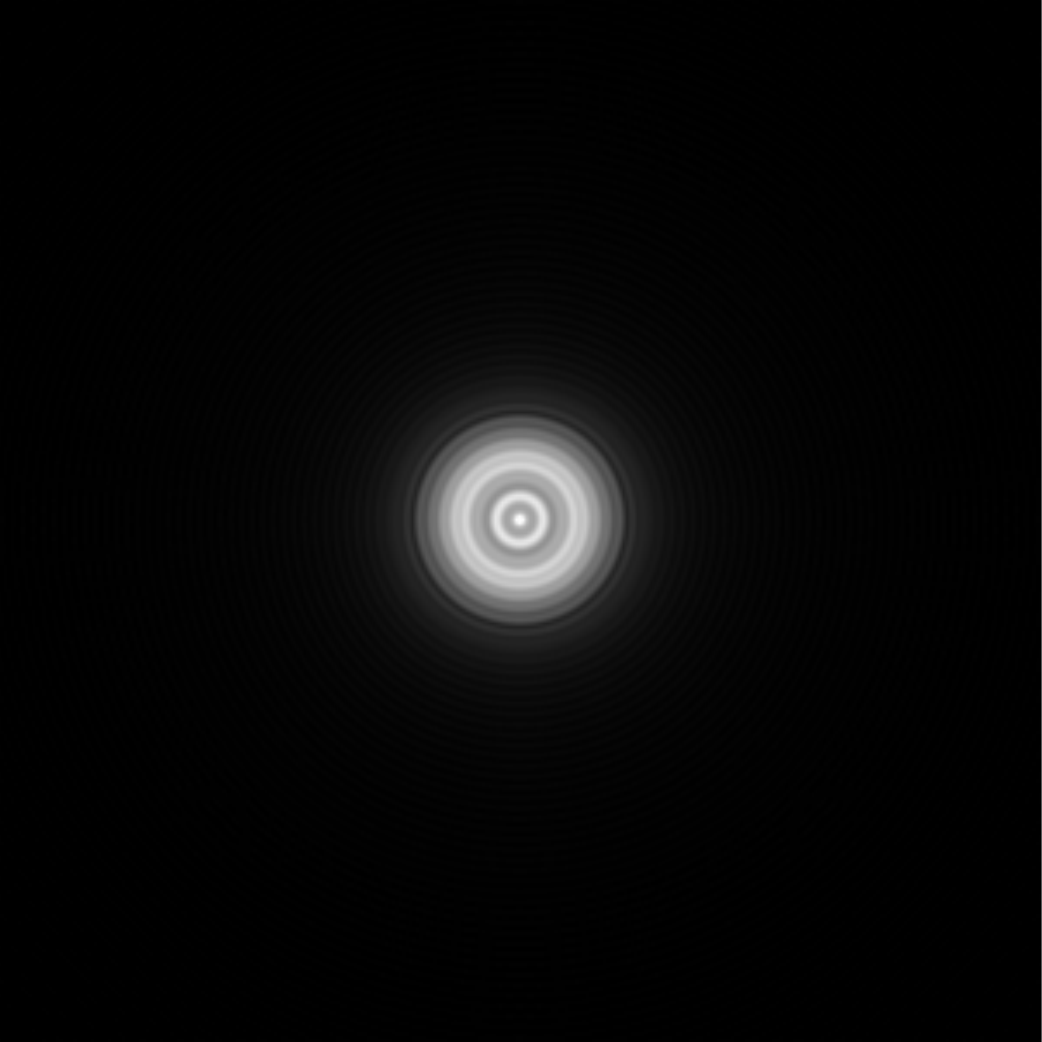}
  \caption{Images of black holes reconstructed from scattering waves
    ($M\omega=12, r_\text{S}=6M$). From the top left panel to the
    bottom right panel, the scattering angles are
    $\theta_0=0,\pi/4,\pi/2,3\pi/4,\pi$.}
\end{figure}

\begin{figure}[H]
  \centering
  \includegraphics[width=0.4\linewidth,clip]{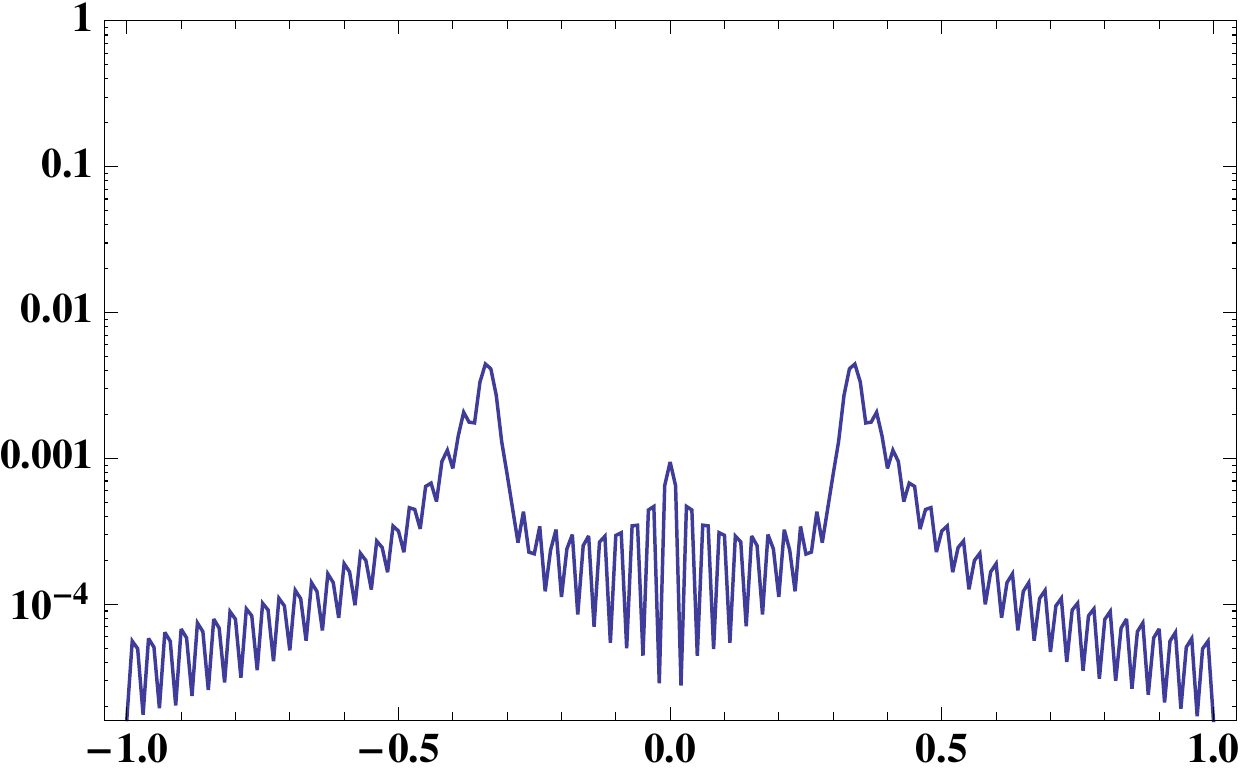}
  \includegraphics[width=0.4\linewidth,clip]{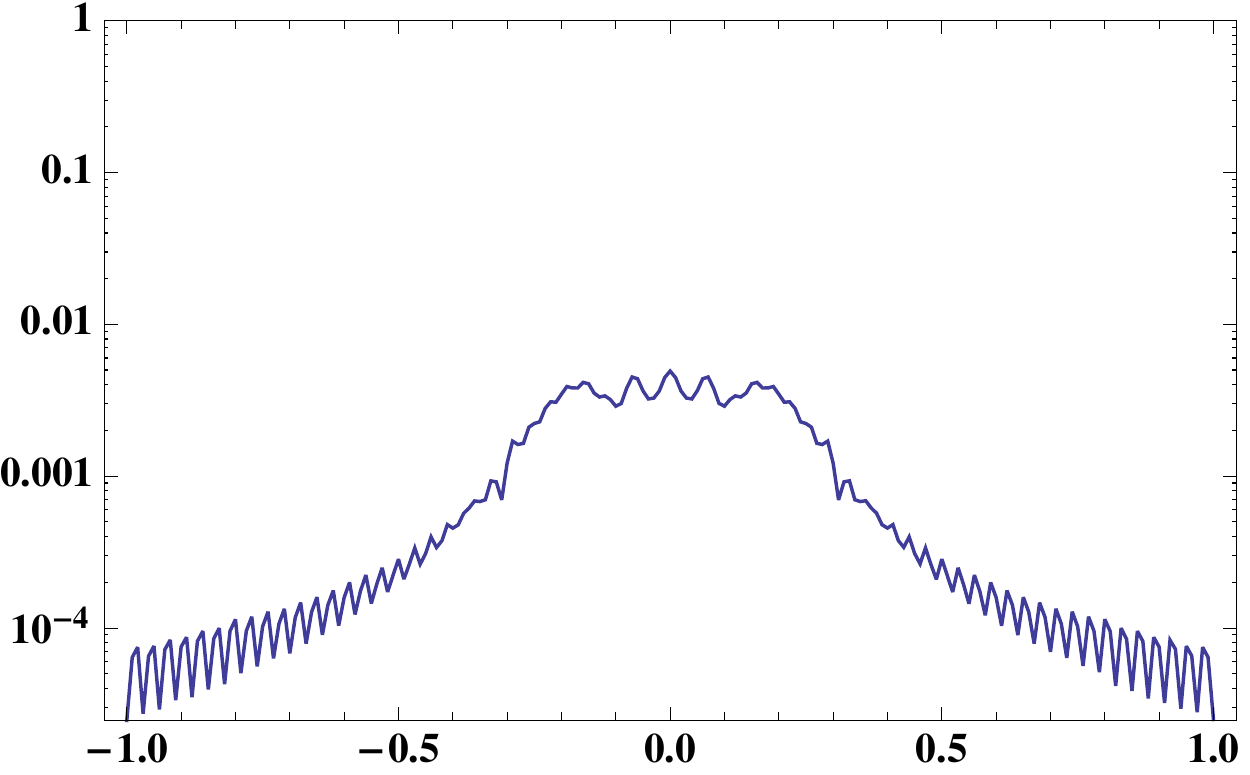}
  \caption{The intensity distribution of images. Left panel:
    $\theta_0=0$ (forward). Right panel: $\theta_0=\pi$ (backward).}
\end{figure}

In the geometric optics limit, images of black holes can be obtained
by solving null geodesics. For the observer at $\theta_0=0$, the
primary null rays, which are deflected by the black hole but do not go
around it, result in the Einstein ring. The secondary and the
higher degrees of null rays that go around the black hole many times
also form ring images with smaller angular radius compared to the
Einstein ring. The left panel in Fig.~9 shows an example of the
primary and the secondary null rays  connecting the source and the
observer at $\theta_0=0$. The right panel in Fig.~9 shows the apparent
angular radius of the Einstein ring and the ring by the secondary
rays as the function of the source position $r_\text{S}$.  For
$r_\text{S}=6M$, the ratio of angular sizes of these two rings are
$\sim 1.2$.
\begin{figure}[H]
  \centering
  \includegraphics[width=0.4\linewidth,clip]{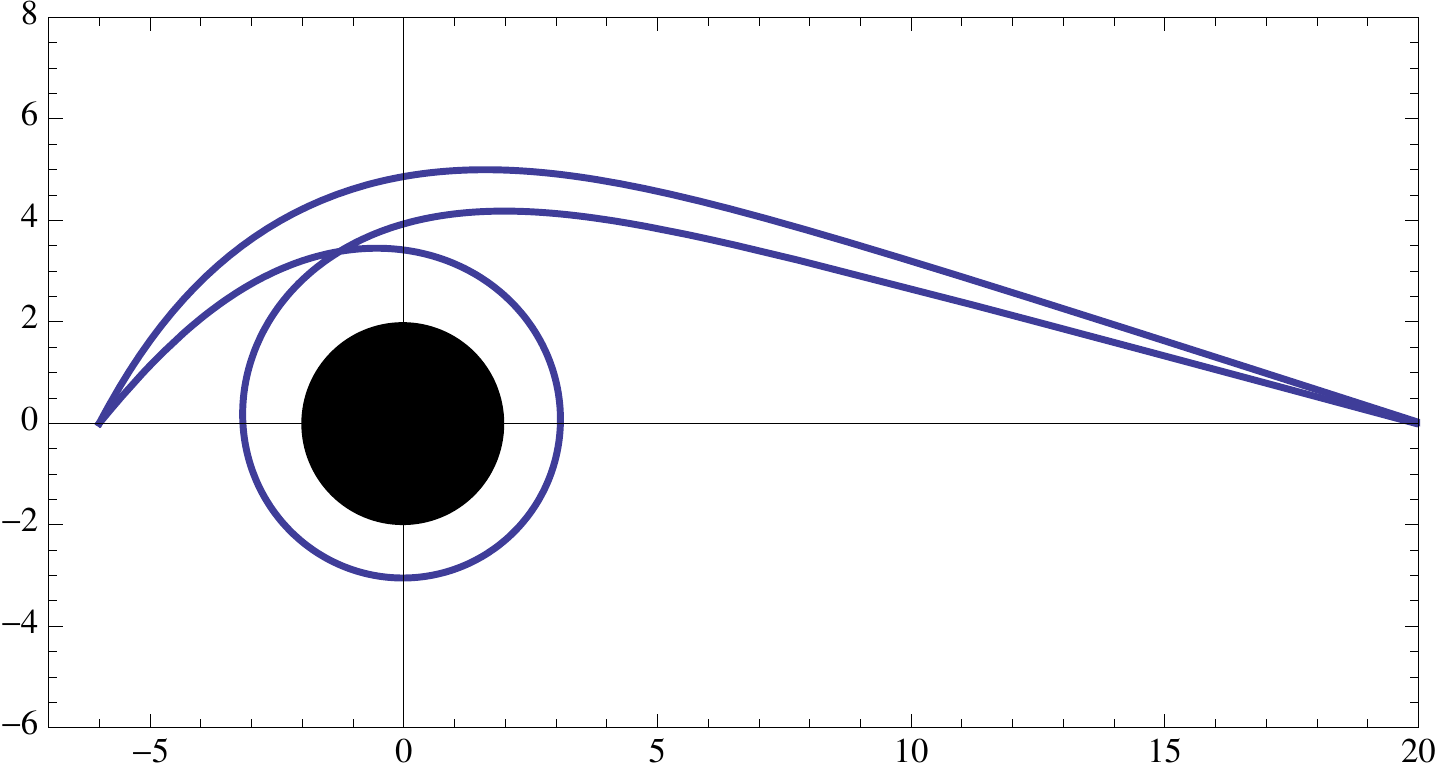}
  \includegraphics[width=0.45\linewidth,clip]{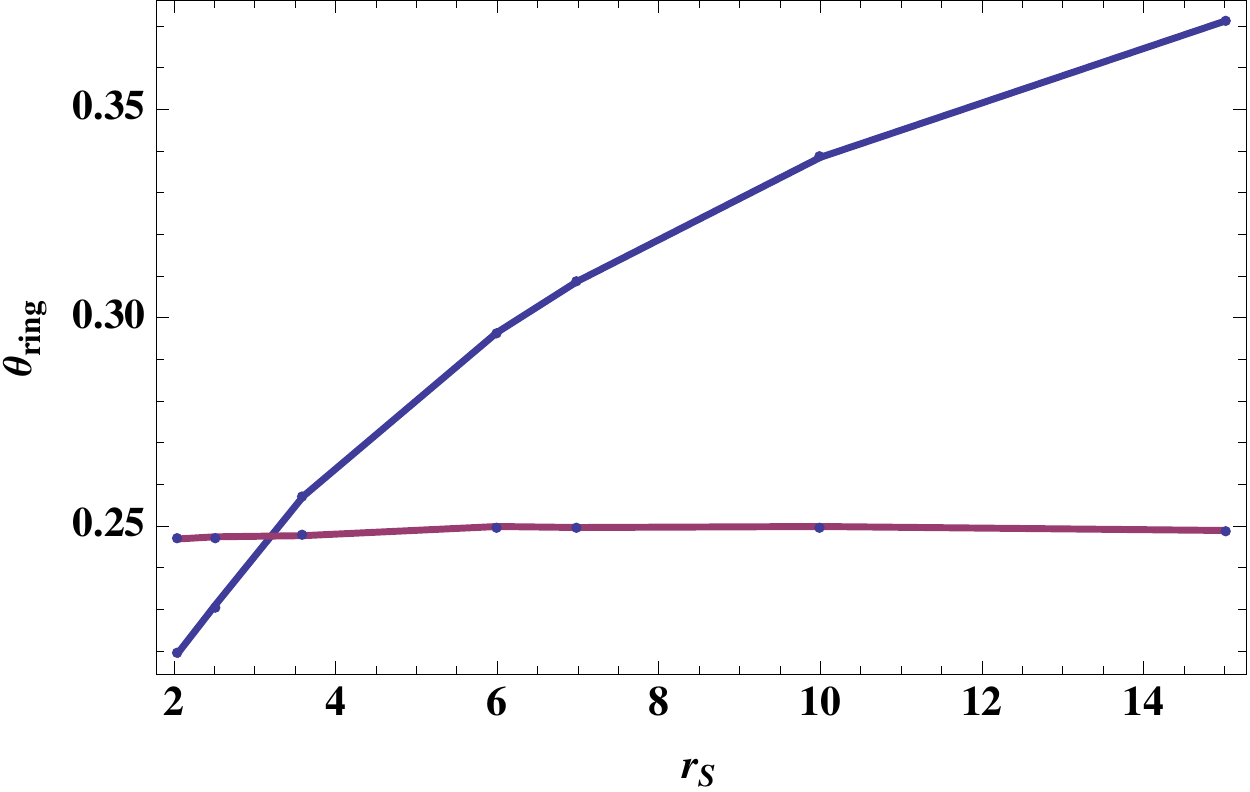}
  \caption{Left panel: the primary and the secondary null rays
    connecting the source at $r_\text{S}=6M$ and the observer at
    $r_\text{obs}=20M$. These rays constitute ring images. Right
    panel: The apparent angular radius of ring images for different
    location of the source. The blue line corresponds to the primary
    rays (Einstein ring) and the red line corresponds to the secondary
    rays.}
\end{figure}
\noindent
We expect to observe these double ring structure in our image
reconstructed from the scattering waves. However, as are shown in
Fig.~7 and Fig.~8, it is not possible to identify this structure
because it is buried in the diffraction pattern appeared in the
image. We cannot conclude the observed second diffraction peak in
Fig.~8 really corresponds to the Einstein ring. By using higher
frequency waves, identification of the double ring structure of the
images will be succeed. For this purpose, we present the image using
the wave with $M\omega=24$ (Fig.~10). The highest peak of the
intensity corresponds to the unstable orbit and it is possible to
observe a broad second peak outside of it. The measured radius of the
second peak is about 1.2 times larger than that of the unstable orbit
and we can conclude that the second peak corresponds to the Einstein
ring.
\begin{figure}[H]
  \centering
  \includegraphics[width=0.35\linewidth,clip]{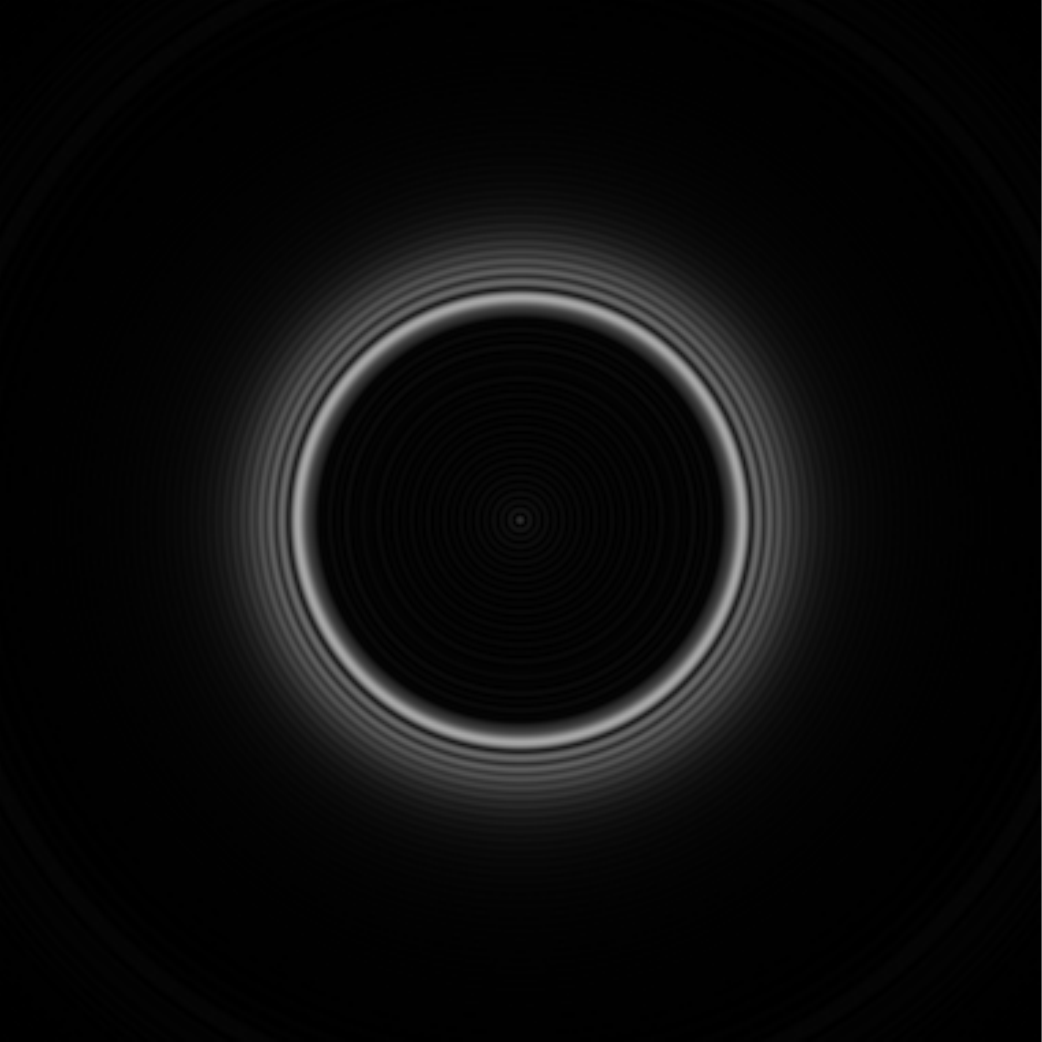}
  \hspace{0.5cm}
  \includegraphics[width=0.45\linewidth,clip]{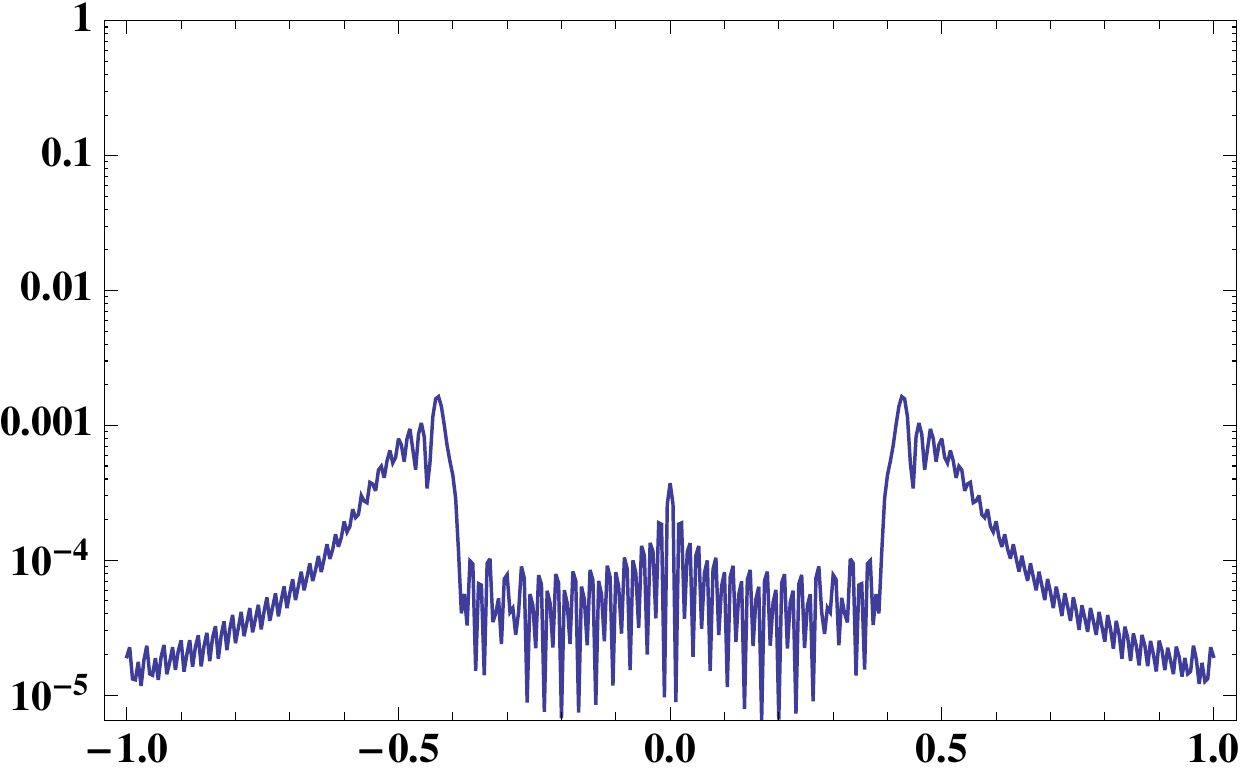}
  \caption{Left panel: image for $\theta_0=0$ using the wave with
    $M\omega=24, d/r_\text{obs}=0.6$. Right panel: intensity
    distribution of the image.}
\end{figure}

 We then consider the case that the source is near the black hole 
$r_\text{S}=2.5M$; in this case, the source is placed inside the unstable
orbit $3M$. We do not observe significant difference of the scattering
behavior of waves compared to $r_\text{S}=6M$ case (Figs.~11 and 12).
\begin{figure}[H]
  \centering
  \includegraphics[width=0.4\linewidth,clip]{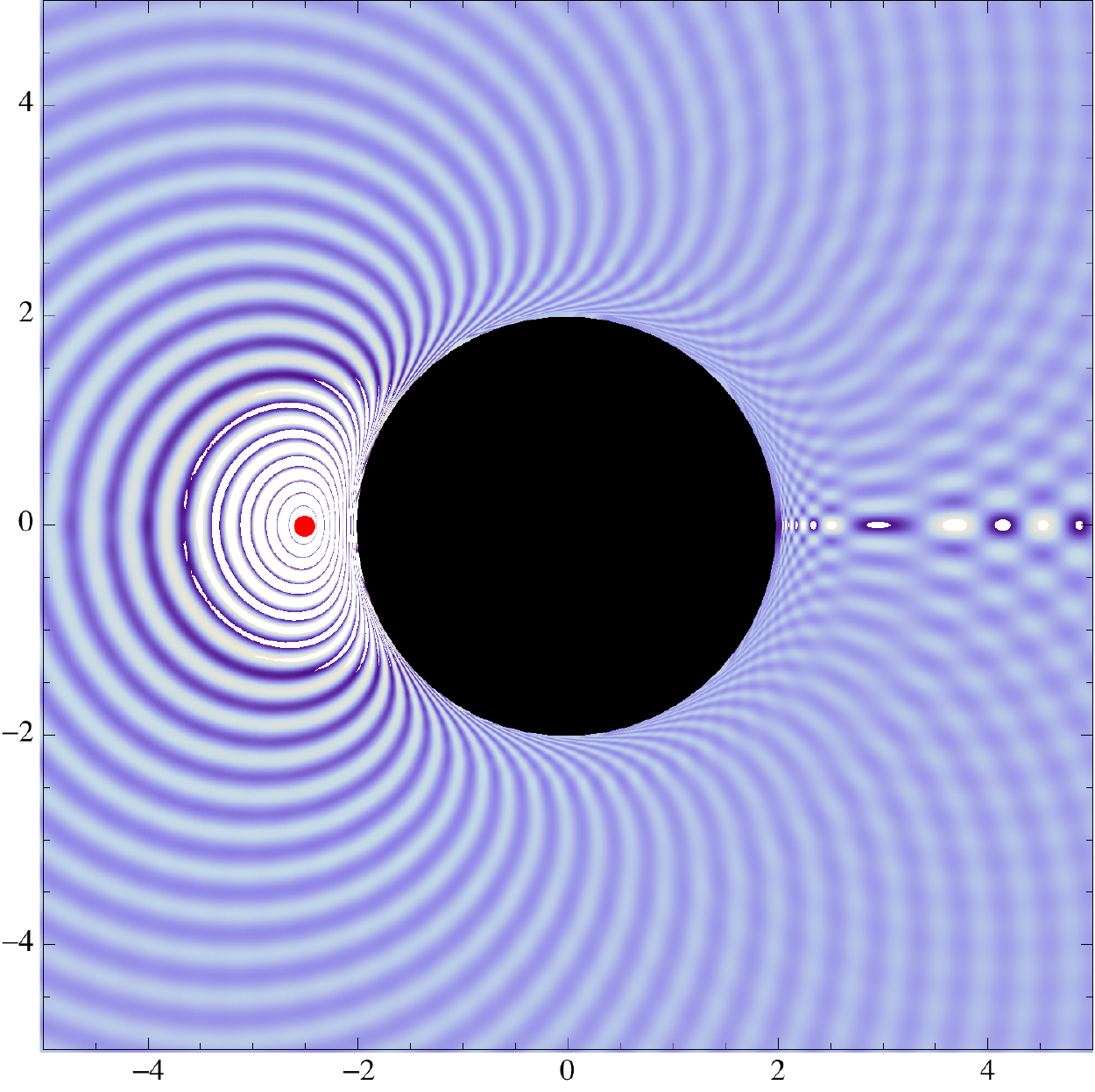}%
  \hspace{0.5cm}
  \includegraphics[width=0.4\linewidth,clip]{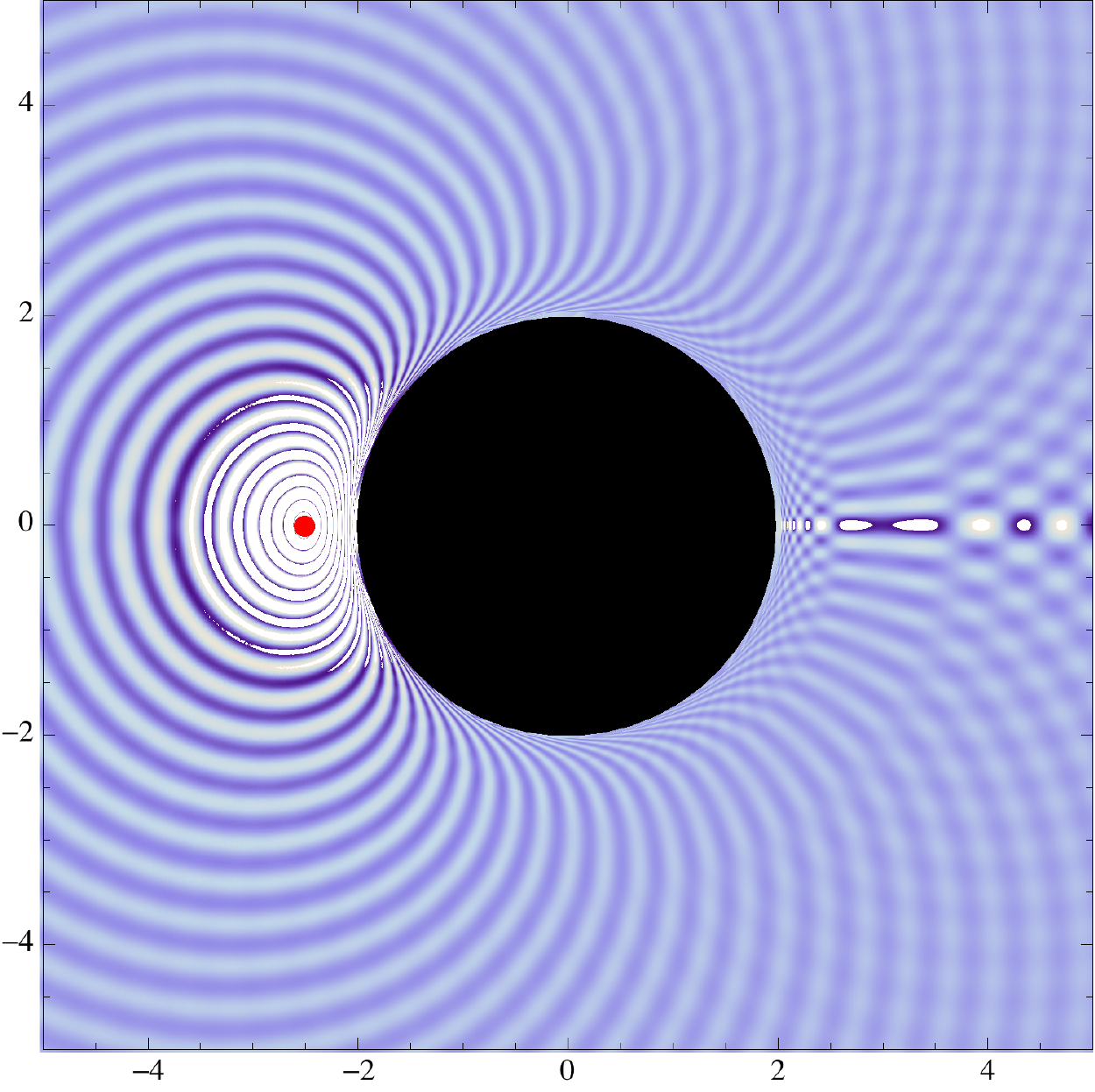}
  \caption{Spatial distribution of $\Phi$ on $(\bar{z},\bar{x})$ plane
    for $M\omega=12$. Left panel: $\text{Re}[\Phi]$. Right panel:
    $\text{Im}[\Phi]$. The point source is located at
    $(\bar{z},\bar{x})=(-2.5M,0)$.}
\end{figure}

\begin{figure}[H]
  \centering
  \includegraphics[width=0.5\linewidth,clip]{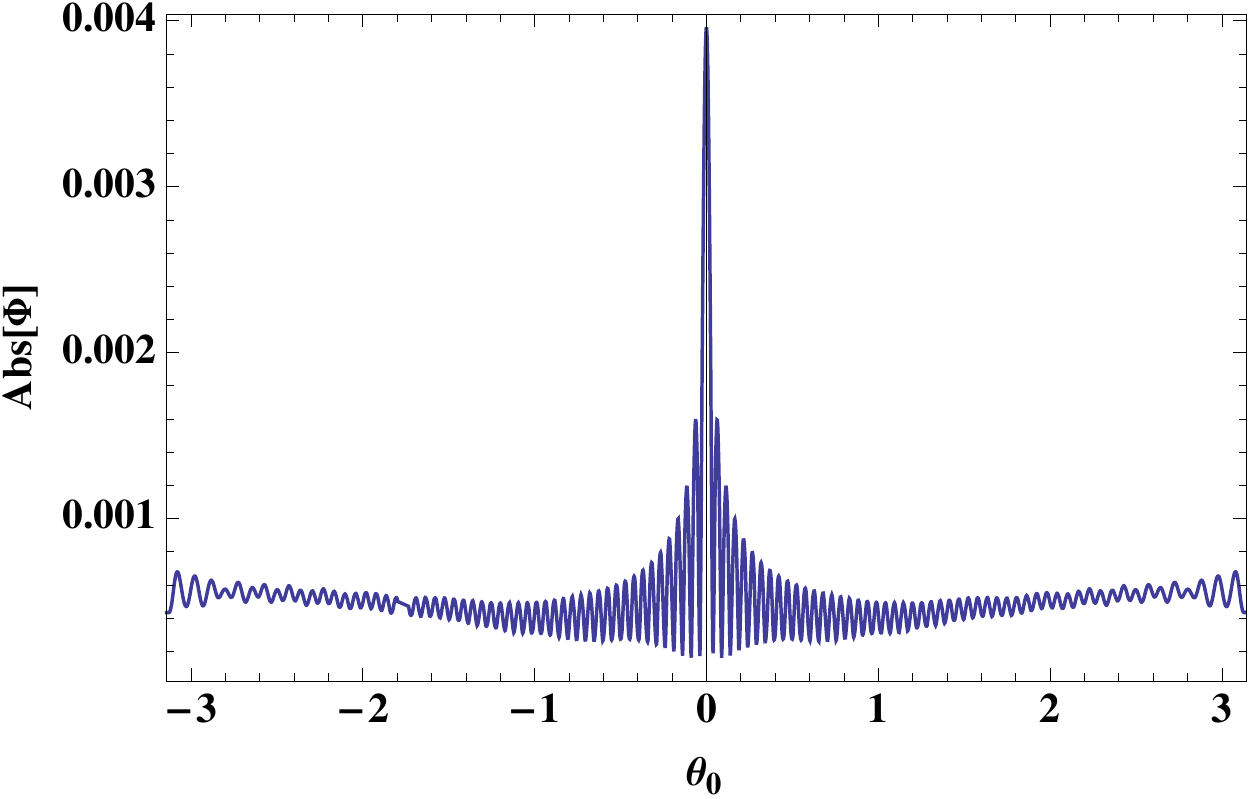}
  \caption{The scattering amplitude at
    $r_\text{obs}=20M$. $M\omega=12, r_\text{S}=2.5M$.}
\end{figure}
In this case, it is possible to observe a ring for the backward
direction $\theta_0=\pi$ in image (Figs.~13 and 14) because the distance
between the source and the observer is larger than $r_\text{S}=6M$
case and the amplitude of the wave that directly reaches the observer
reduces. This enables us to observe the backward glory as a ring
image.
\begin{figure}[H]
  \centering
  \includegraphics[width=0.3\linewidth,clip]{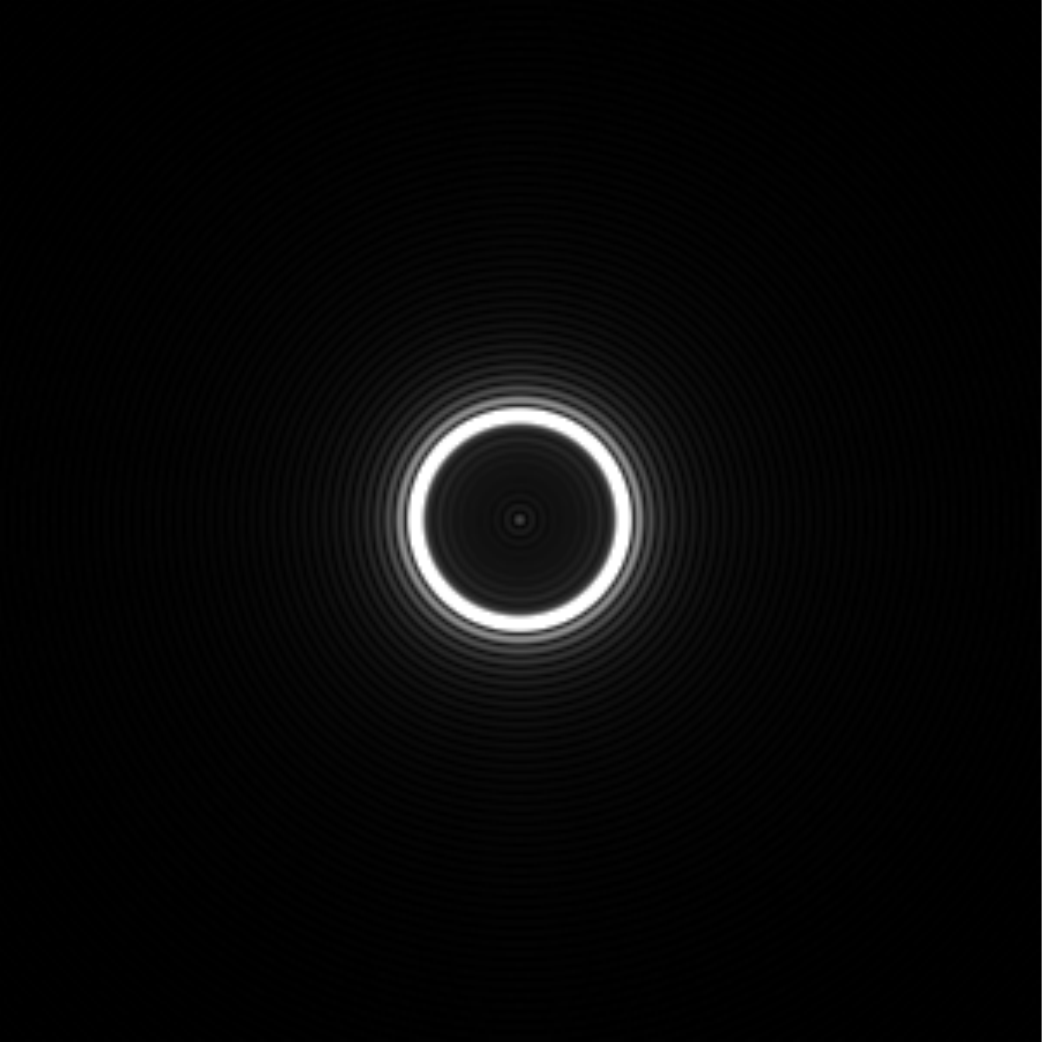}
  \includegraphics[width=0.3\linewidth,clip]{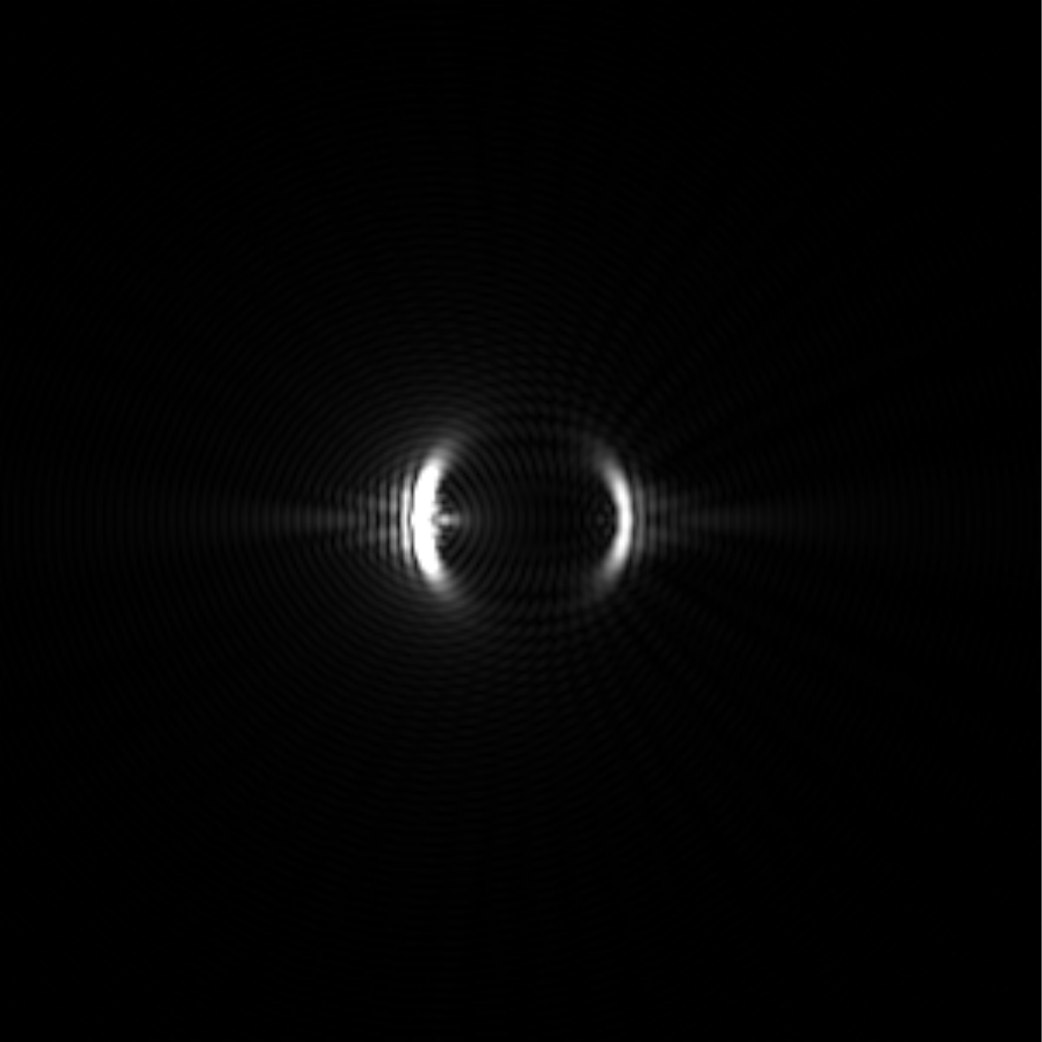}
  \includegraphics[width=0.3\linewidth,clip]{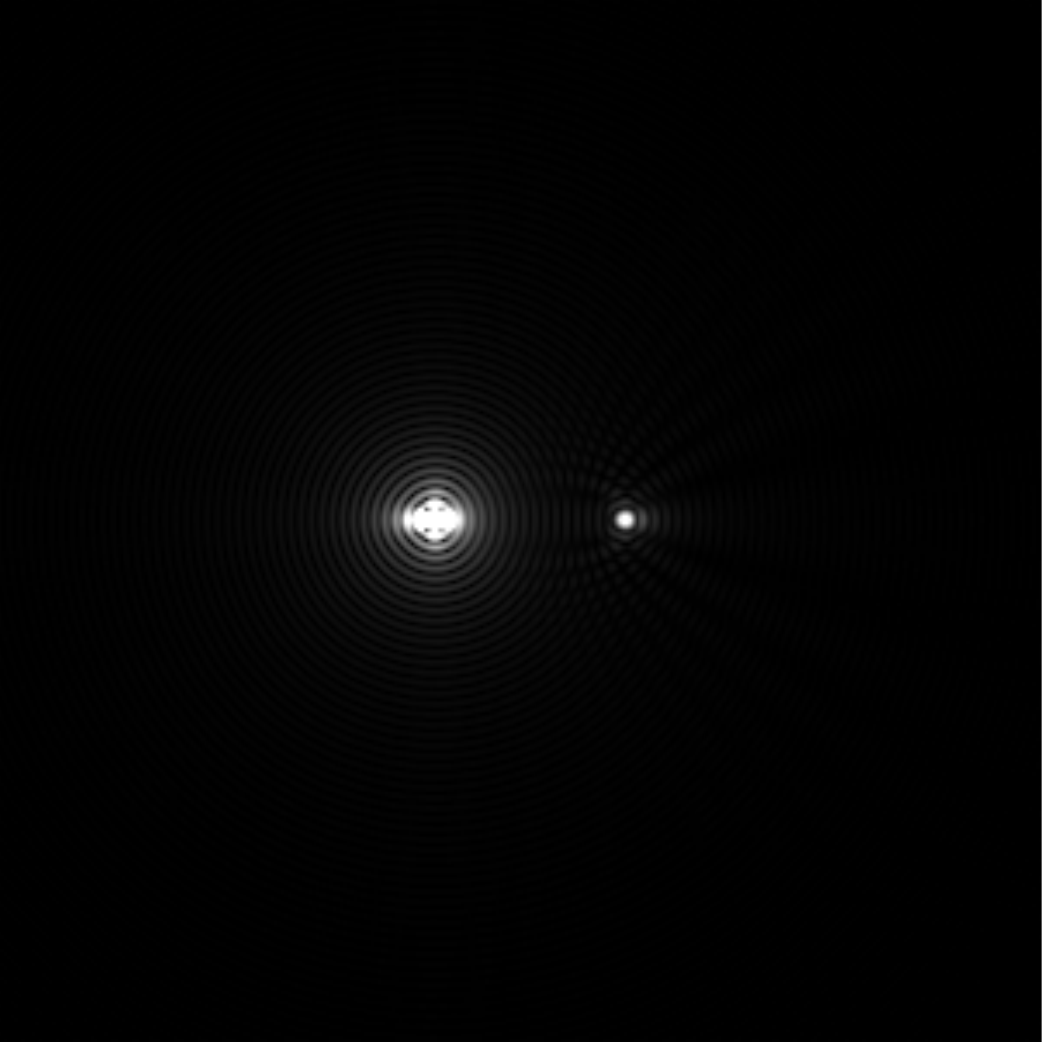}
  \includegraphics[width=0.3\linewidth,clip]{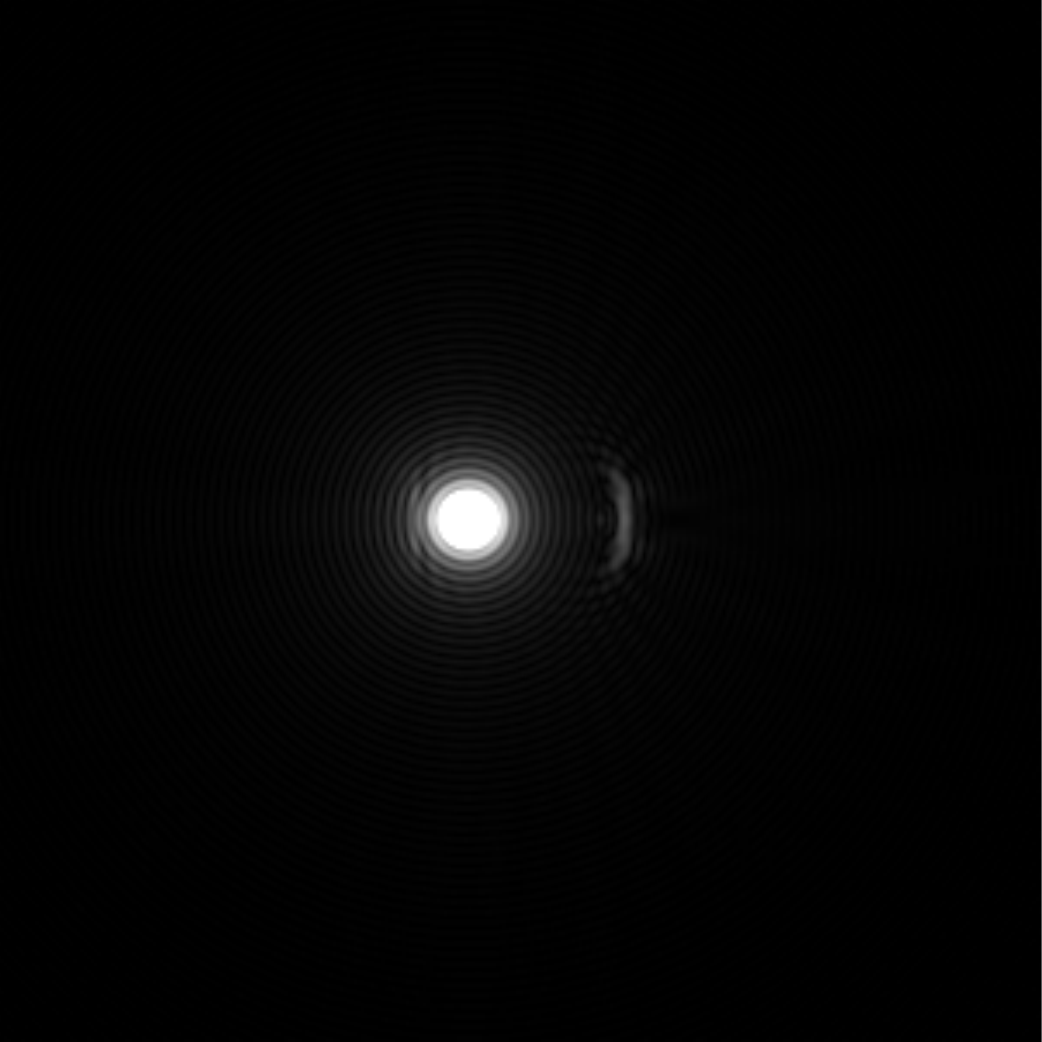}
  \includegraphics[width=0.3\linewidth,clip]{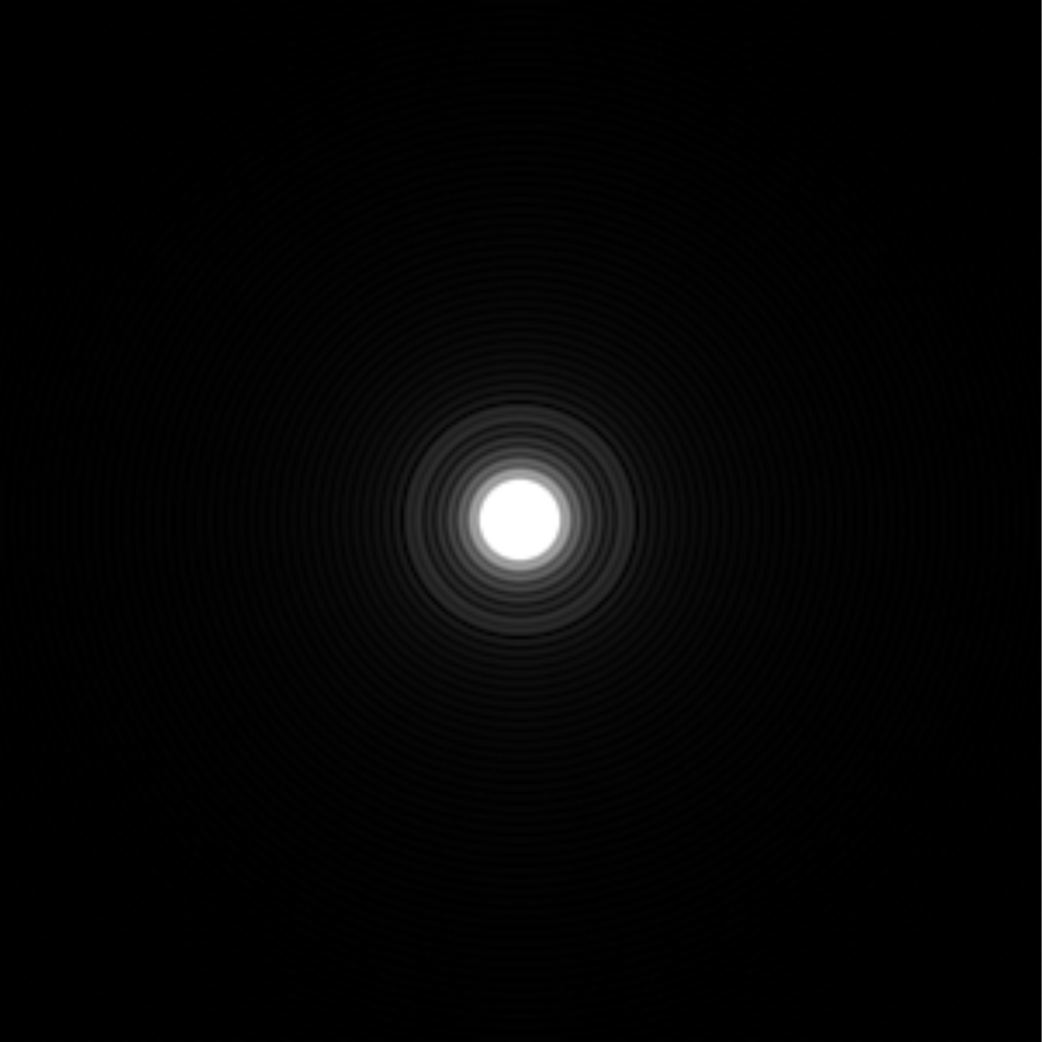}
  \caption{Images of black holes reconstructed from scattering
    waves. From the top left panel to the bottom right panel, the
     scattering angles are $\theta_0=0,\pi/4,\pi/2,3\pi/4,\pi$.}
\end{figure}

\begin{figure}[H]
  \centering
  \includegraphics[width=0.4\linewidth,clip]{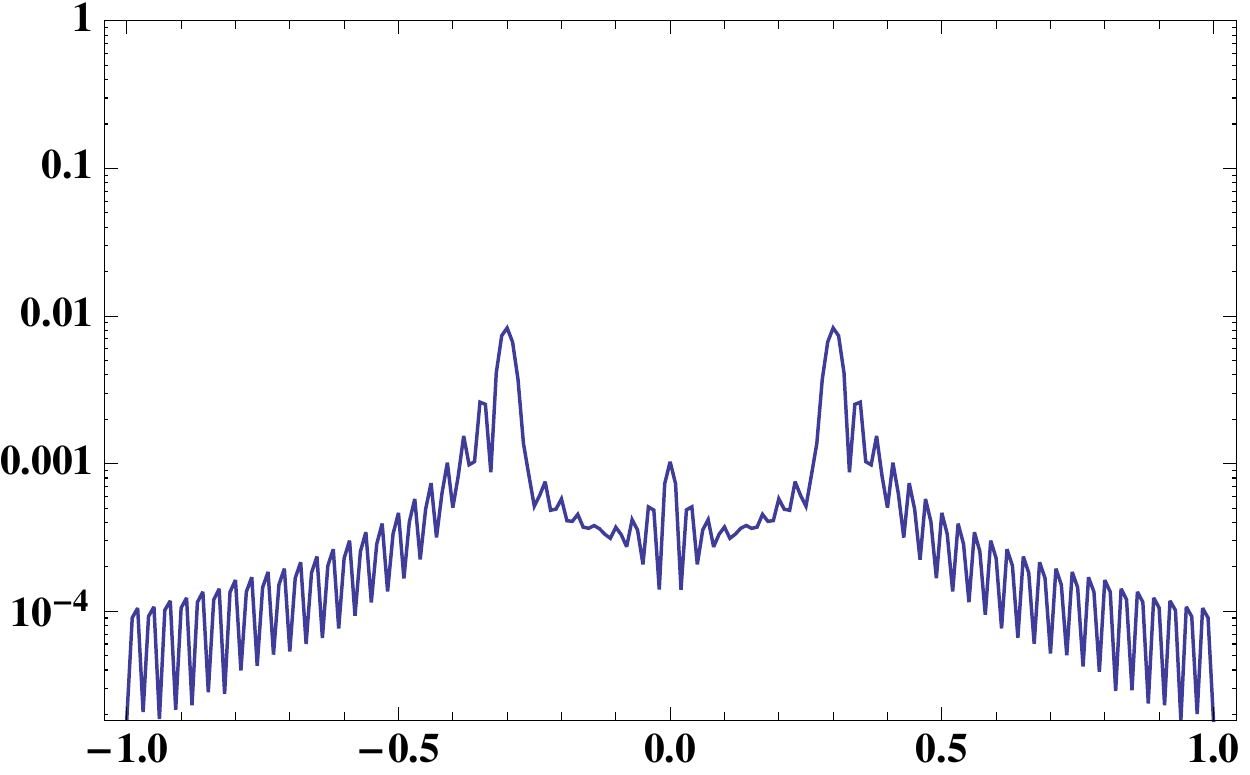}
  \includegraphics[width=0.4\linewidth,clip]{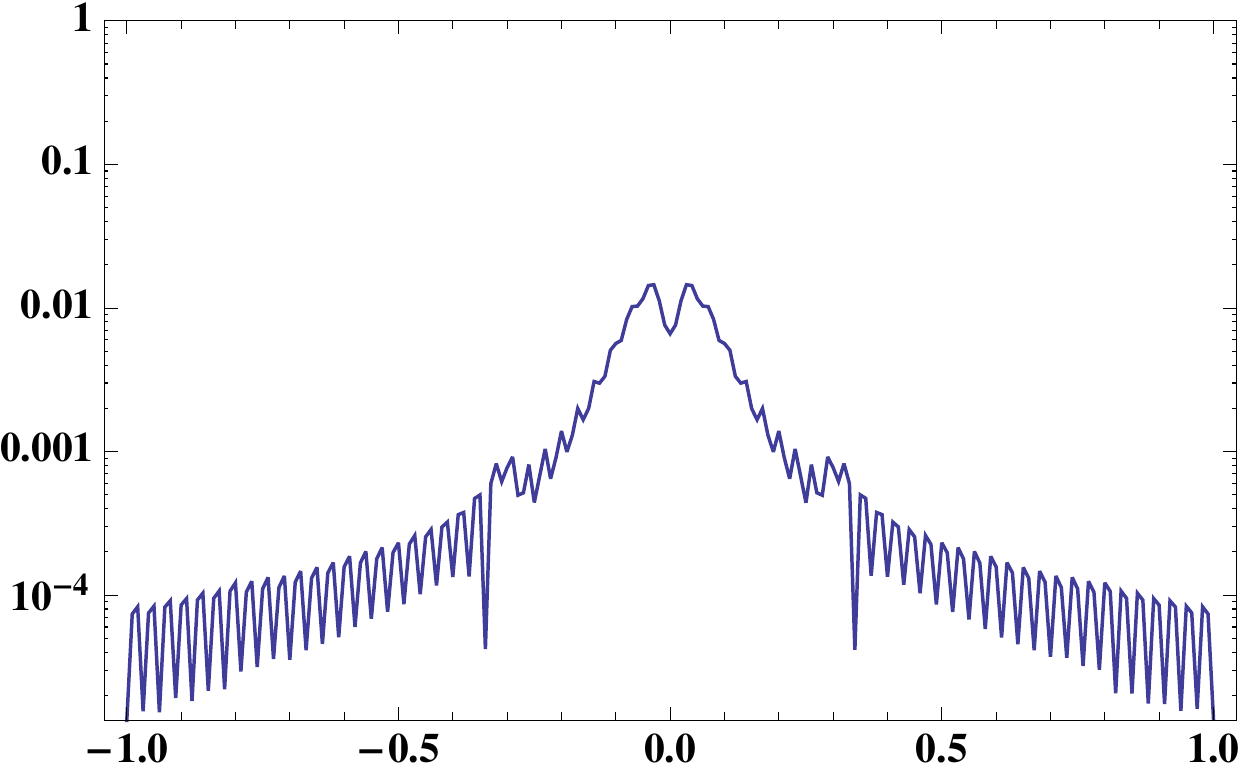}
  \caption{The intensity distribution of images. Left panel:
    $\theta_0=0$. Right panel: $\theta_0=\pi$.}
\end{figure}

\subsection{$M\omega=2$ case}
The wavelength of the source field longer than $M\omega=12$ case and
the wave effects become more significant. Fig.~15 shows spatial distribution
of waves around the black hole. The source position is
$r_\text{S}=6M$.
\begin{figure}[H]
  \centering
  \includegraphics[width=0.4\linewidth,clip]{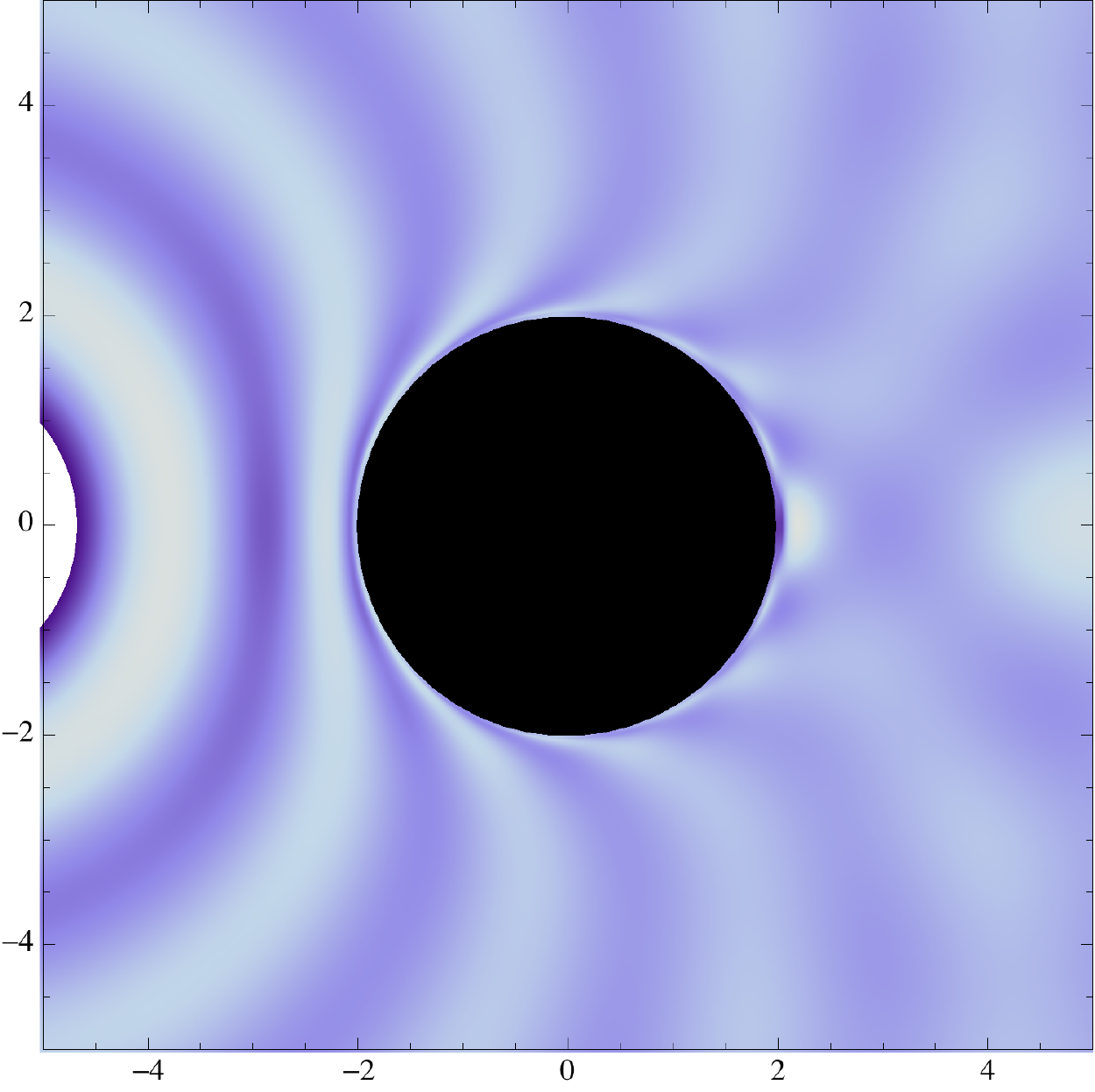}%
  \hspace{0.5cm}
  \includegraphics[width=0.4\linewidth,clip]{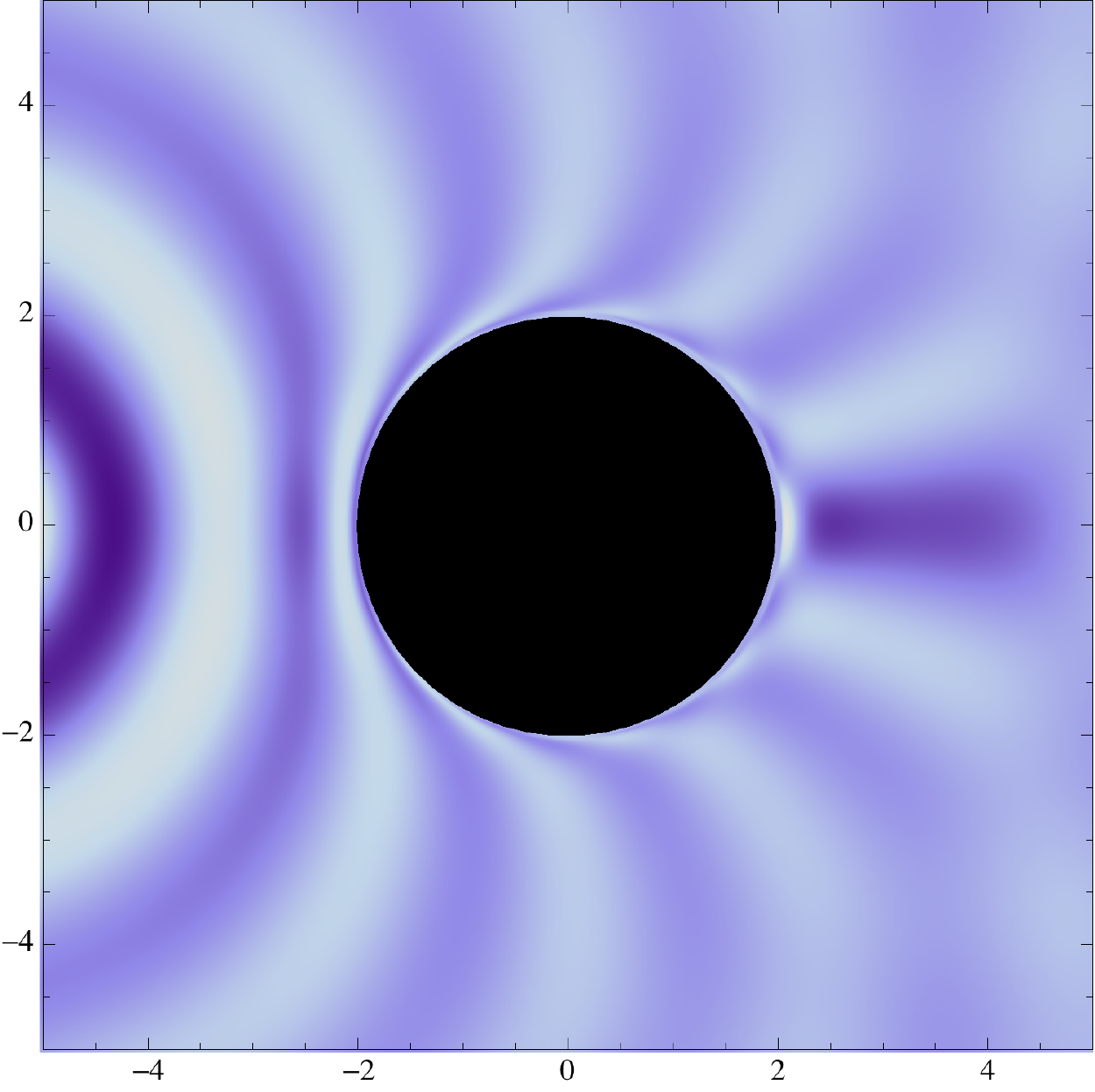}
  \caption{Spatial distribution of $\Phi$ on $(\bar{z},\bar{x})$ plane
    for $M\omega=2$. Left panel: $\text{Re}[\Phi]$. Right panel:
    $\text{Im}[\Phi]$. The point source is located at
    $(\bar{z},\bar{x})=(-6M,0)$.}
\end{figure}
\noindent
The scattering amplitude also shows interference pattern (Fig.~16). The
distance between adjacent fringes becomes larger compared to
$M\omega=12$ case. But the characteristic feature is the same.
\begin{figure}[H]
  \centering
  \includegraphics[width=0.5\linewidth,clip]{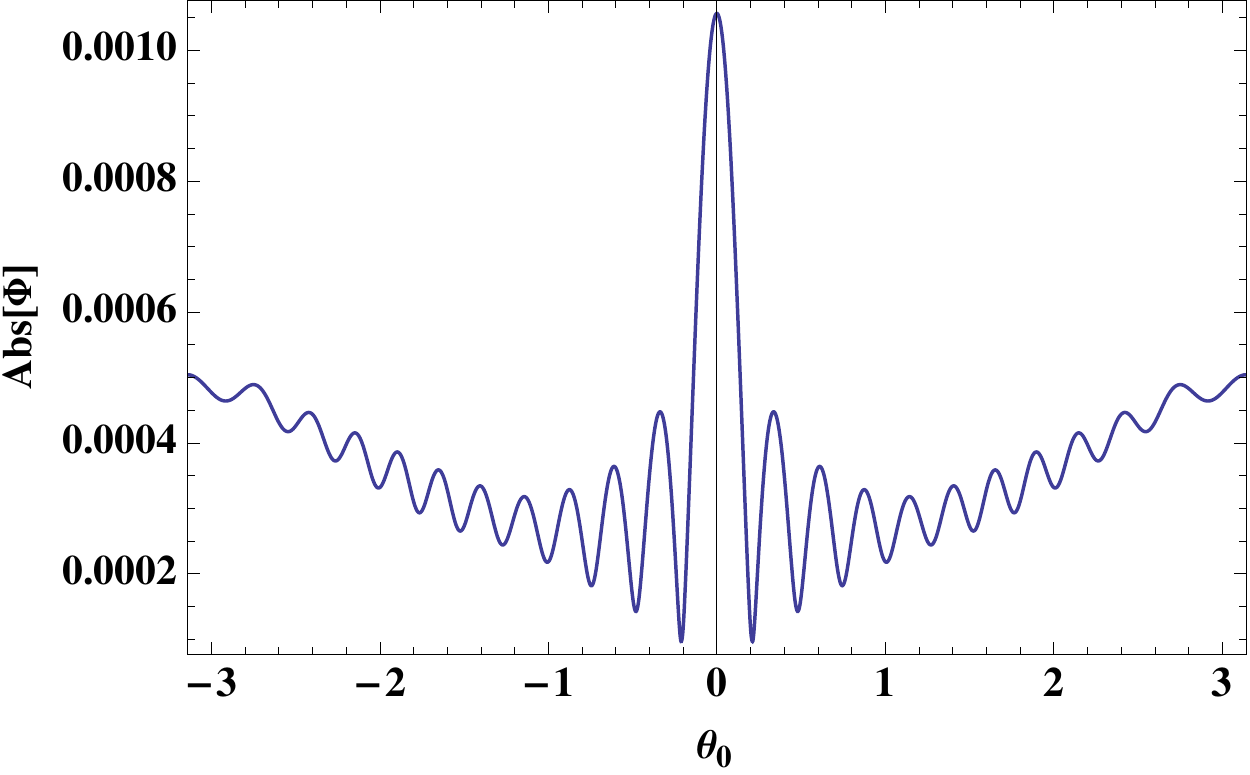}
  \caption{The scattering amplitude at $r_\text{obs}=20M$. $M\omega=2,
  r_\text{S}=6M$.}
\end{figure}
\noindent
In this case, we can also observe a ring image for $\theta_0=0$
corresponding to the unstable orbit (Figs.~17 and 18).
\begin{figure}[H]
  \centering
  \includegraphics[width=0.3\linewidth,clip]{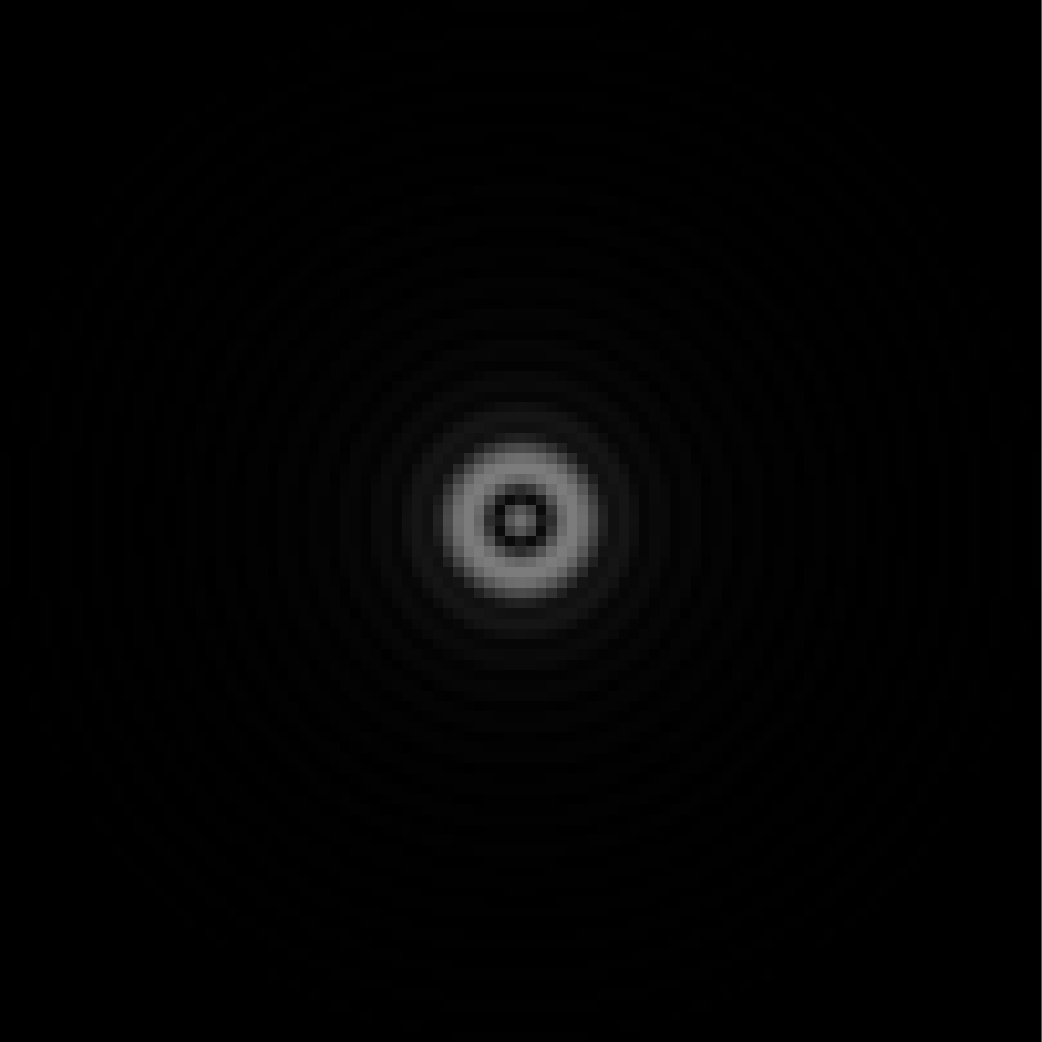}
  \includegraphics[width=0.3\linewidth,clip]{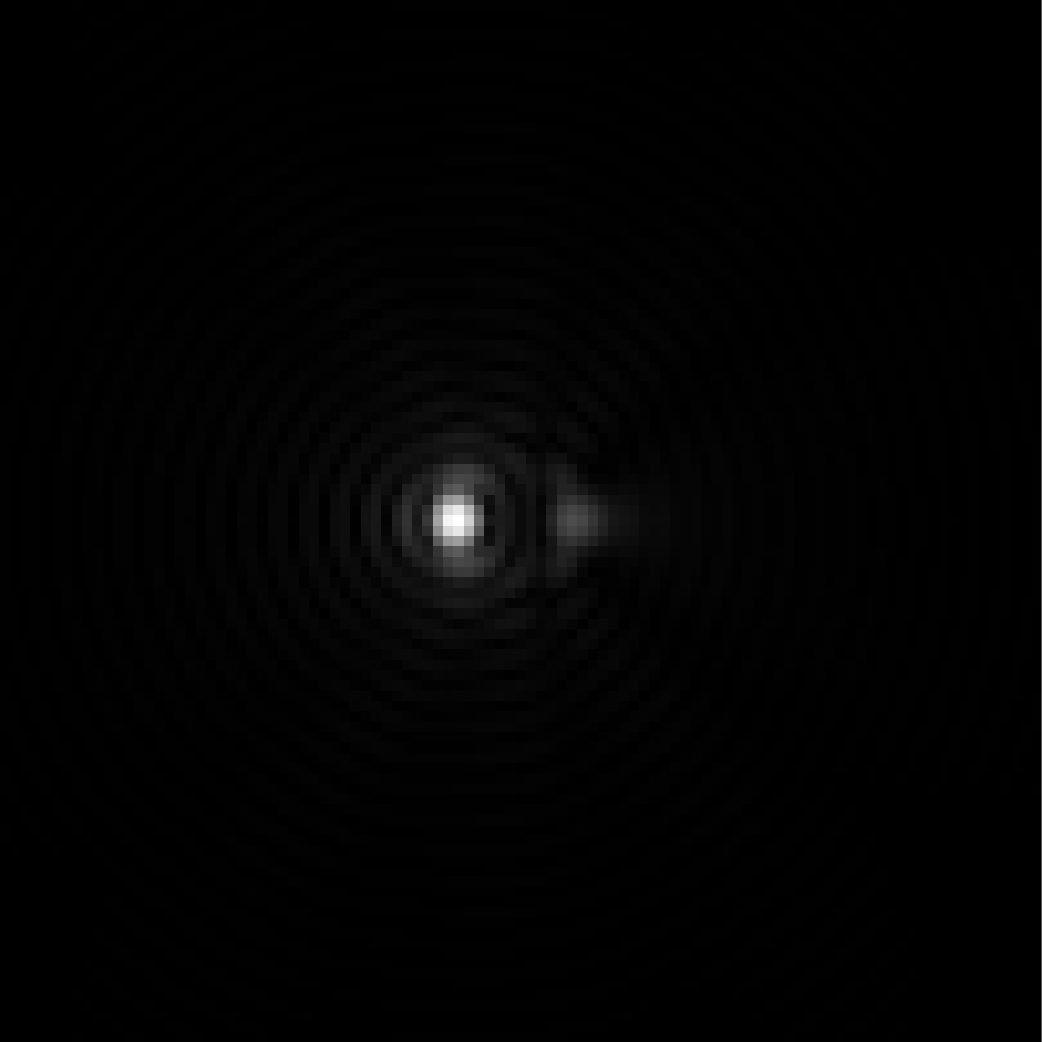}
  \includegraphics[width=0.3\linewidth,clip]{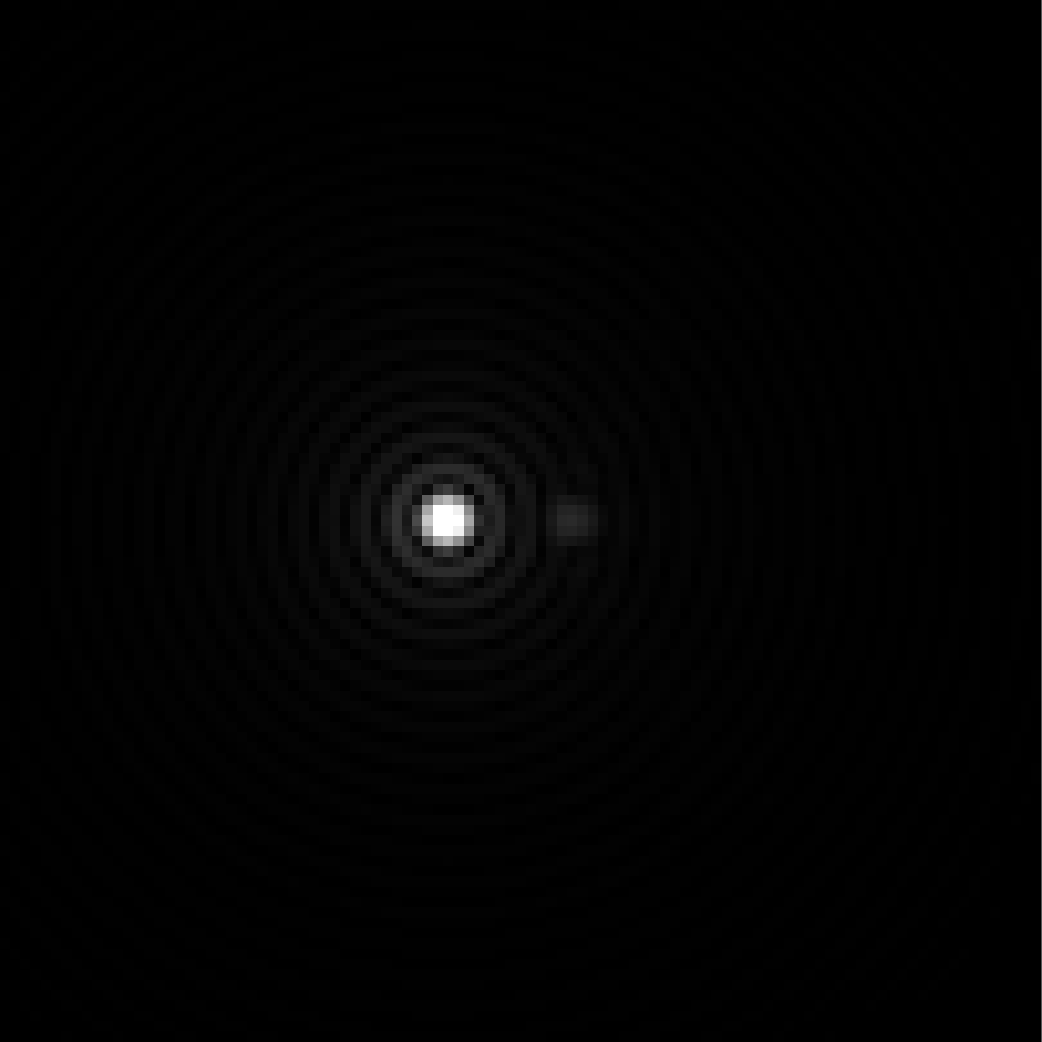}
  \includegraphics[width=0.3\linewidth,clip]{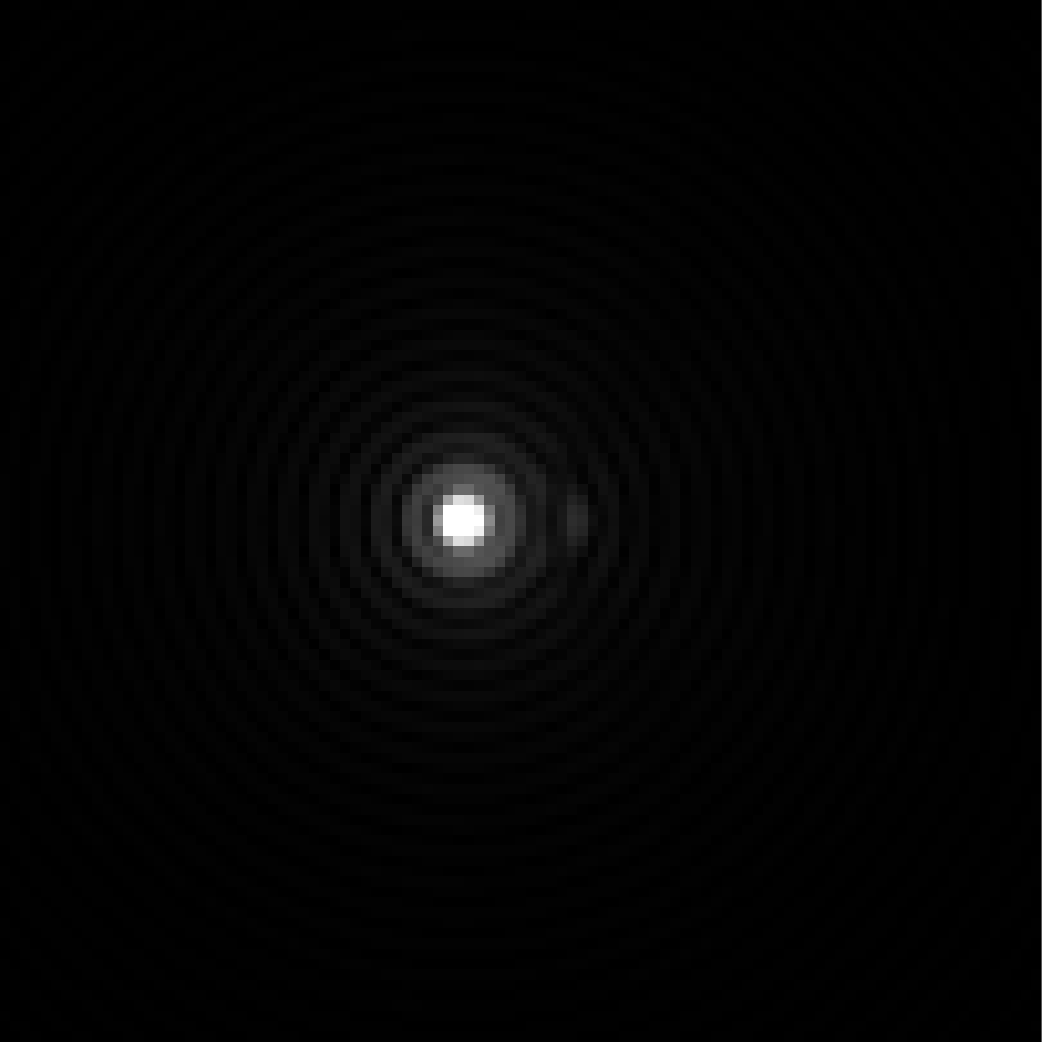}
  \includegraphics[width=0.3\linewidth,clip]{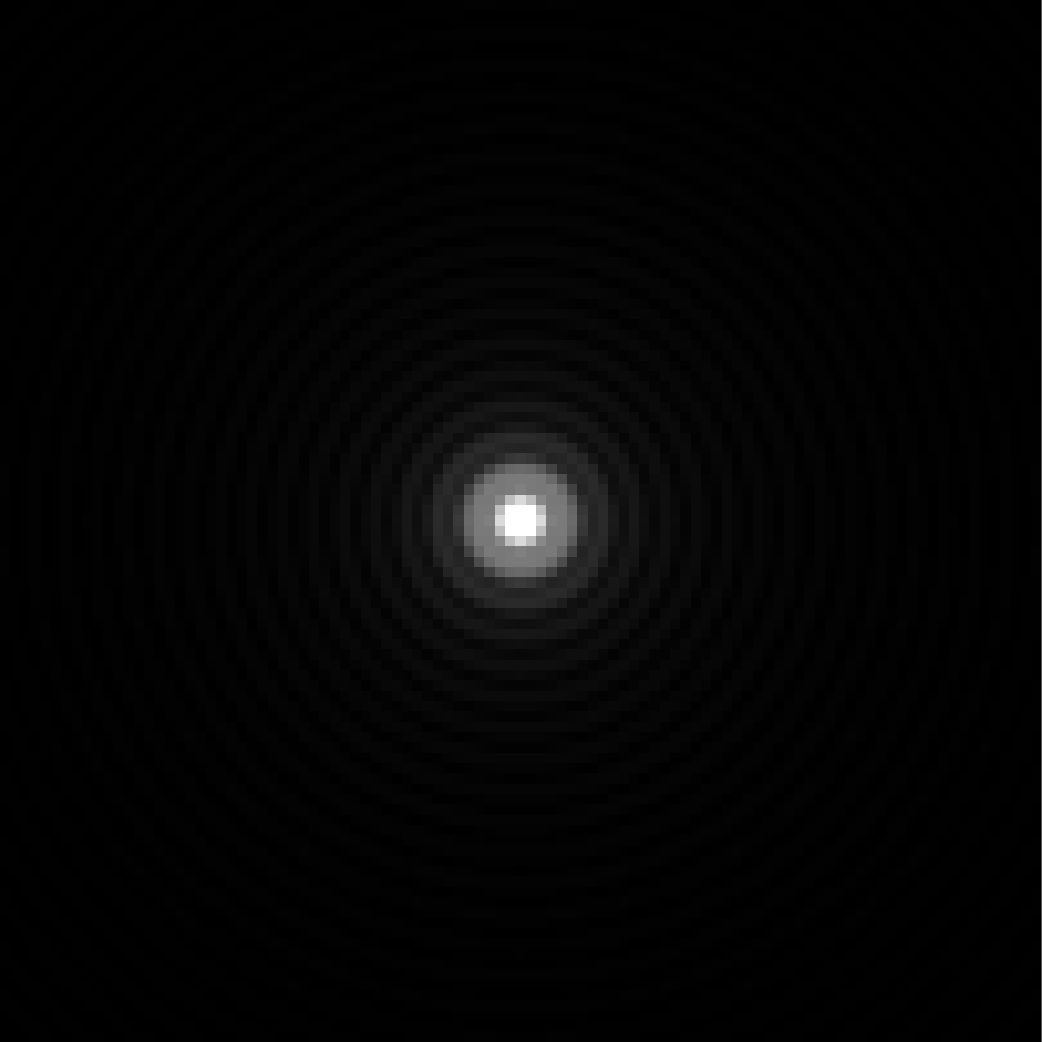}
  \caption{Images of black holes reconstructed from scattering
    waves. From the top left panel to the bottom right panel, the
    scattering angles are $\theta_0=0,\pi/4,\pi/2,3\pi/4,\pi$.}
\end{figure}

\begin{figure}[H]
  \centering
  \includegraphics[width=0.4\linewidth,clip]{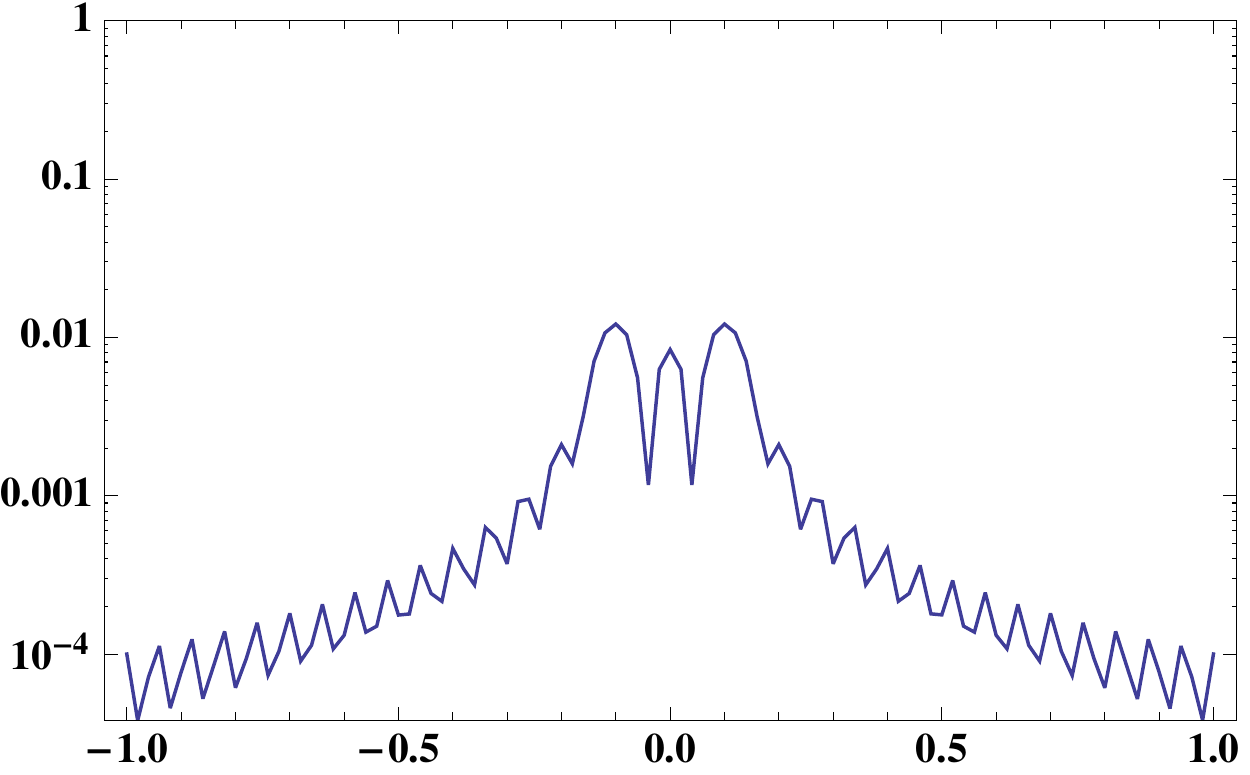}
  \includegraphics[width=0.4\linewidth,clip]{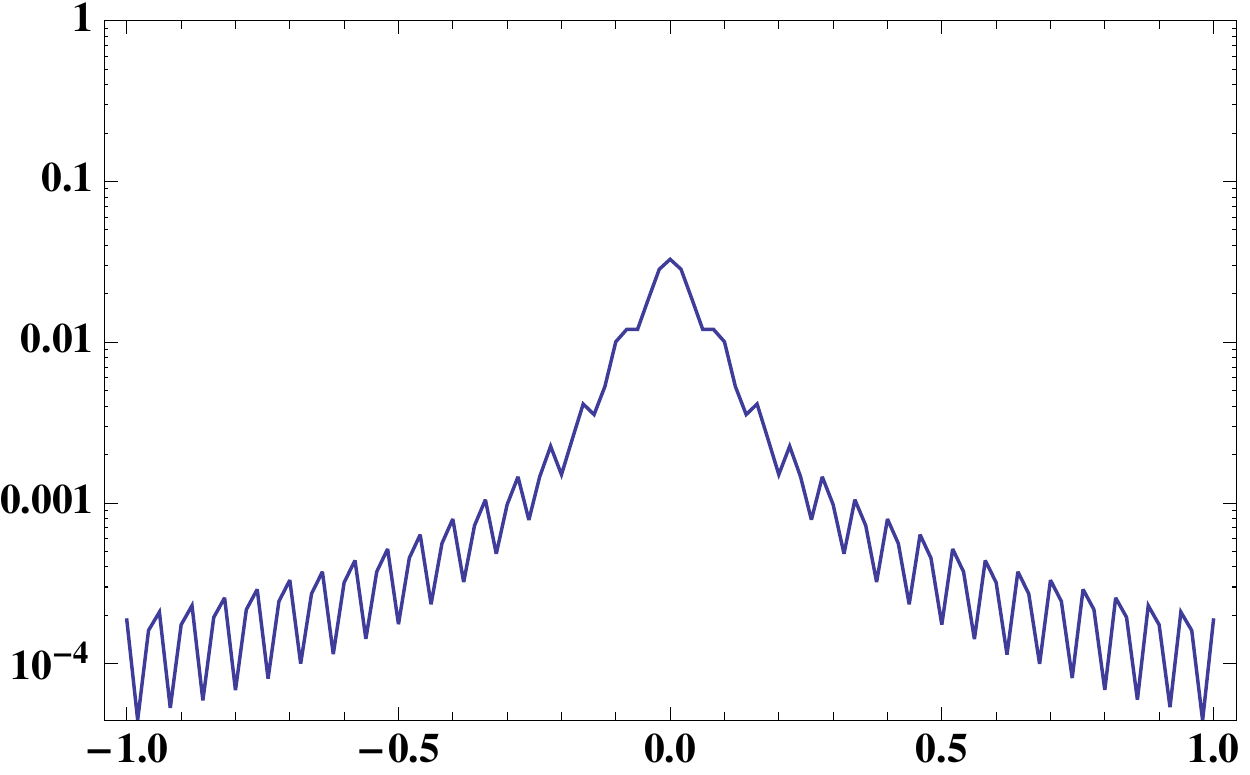}
  \caption{The intensity distribution of images. Left panel:
    $\theta_0=0$. Right panel: $\theta_0=\pi$.}
\end{figure}

Finally, we present the result for the source position
$r_\text{S}=2.5M$ (Figs.~19 and 20). In this case, we can identify
both the forward and the backward glories as ring images (Figs.~21 and
22).
\begin{figure}[H]
  \centering
  \includegraphics[width=0.4\linewidth,clip]{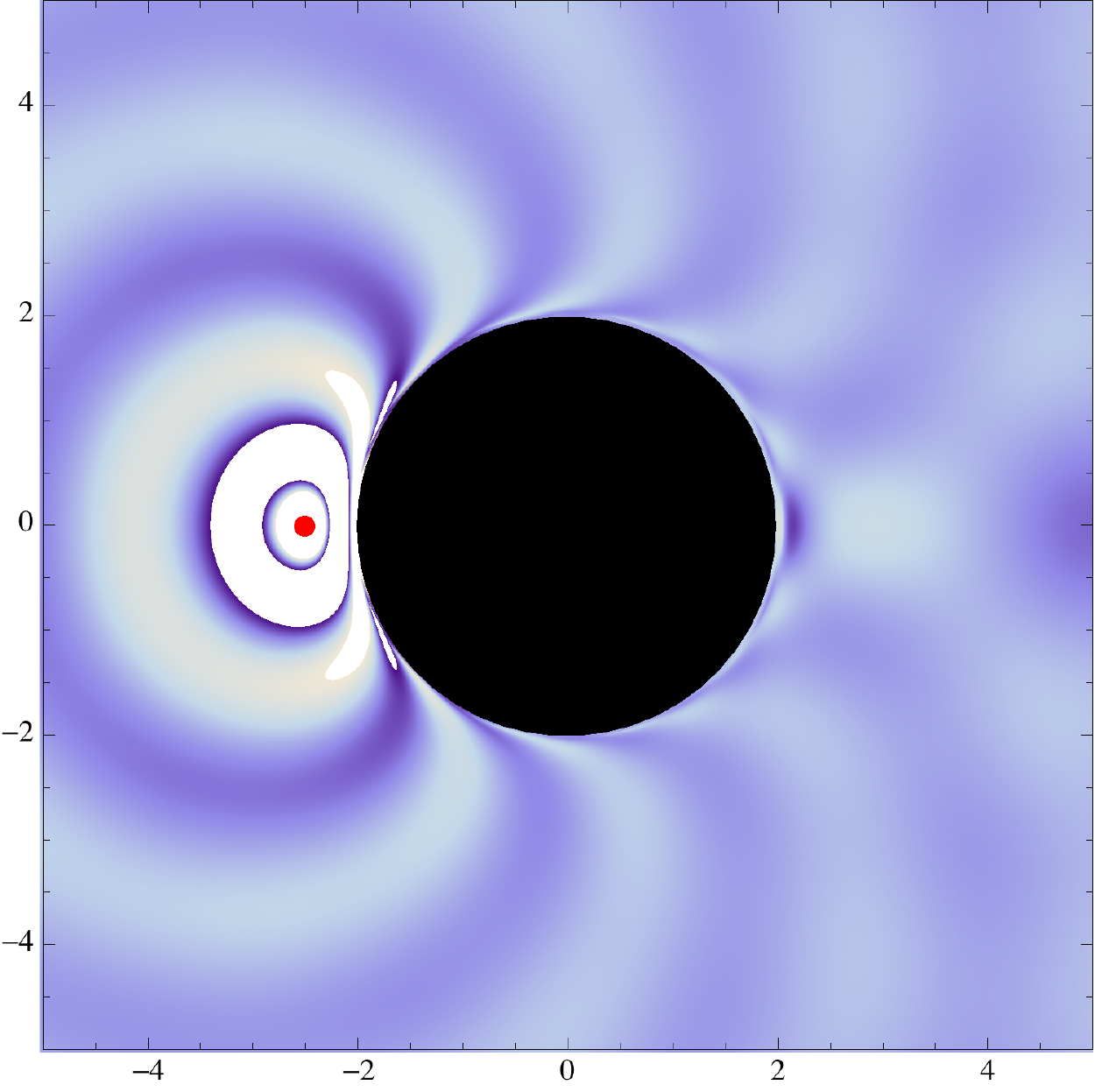}%
  \hspace{0.5cm}
  \includegraphics[width=0.4\linewidth,clip]{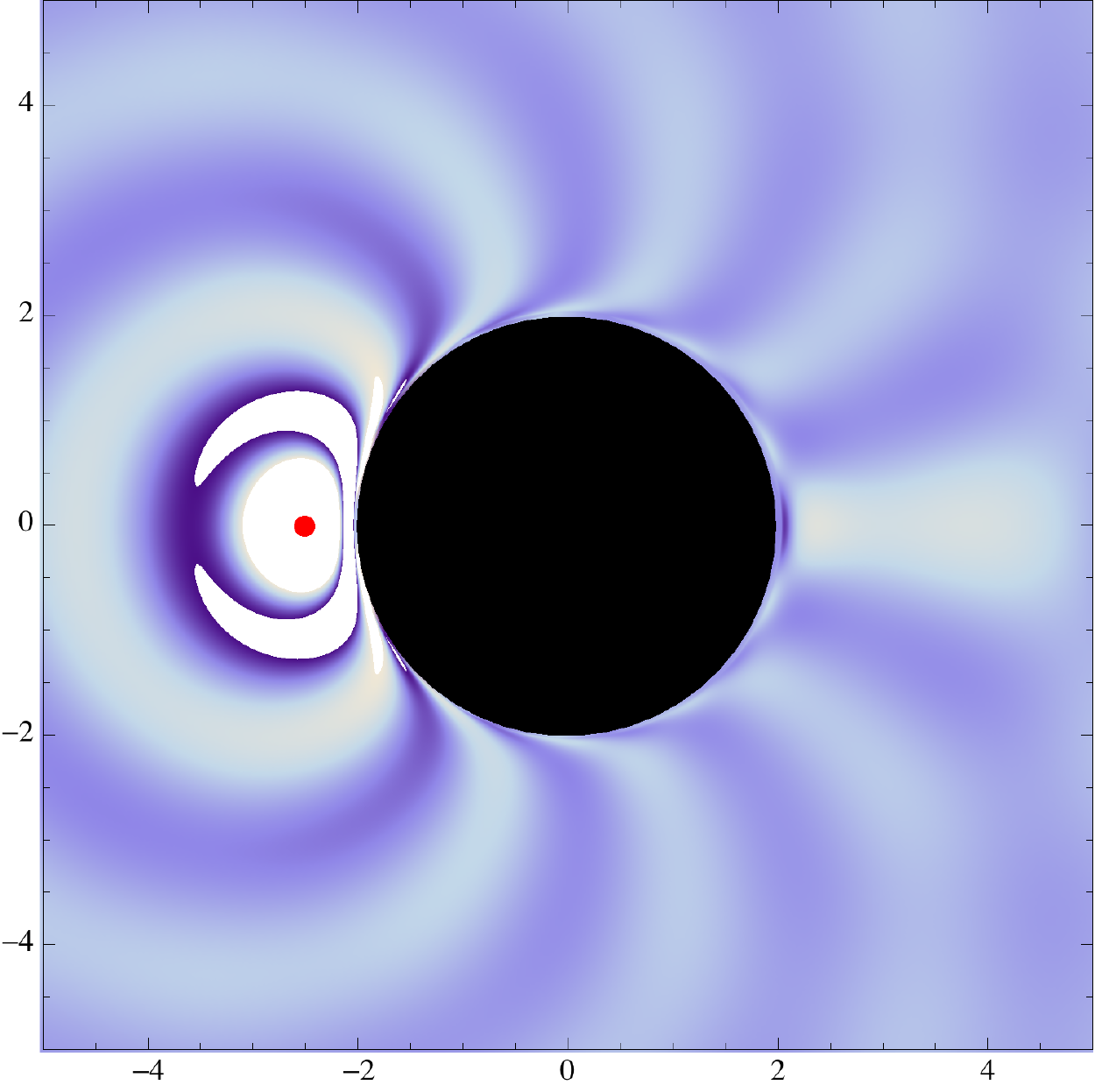}
  \caption{Spatial distribution of $\Phi$ on $(\bar{z},\bar{x})$ plane
    for $M\omega=2$. Left panel:
    $\text{Re}[\Phi]$. Right panel: $\text{Im}[\Phi]$. The point source is
    located at $(\bar{z},\bar{x})=(-2.5M,0)$.}
\end{figure}

\begin{figure}[H]
  \centering
  \includegraphics[width=0.5\linewidth,clip]{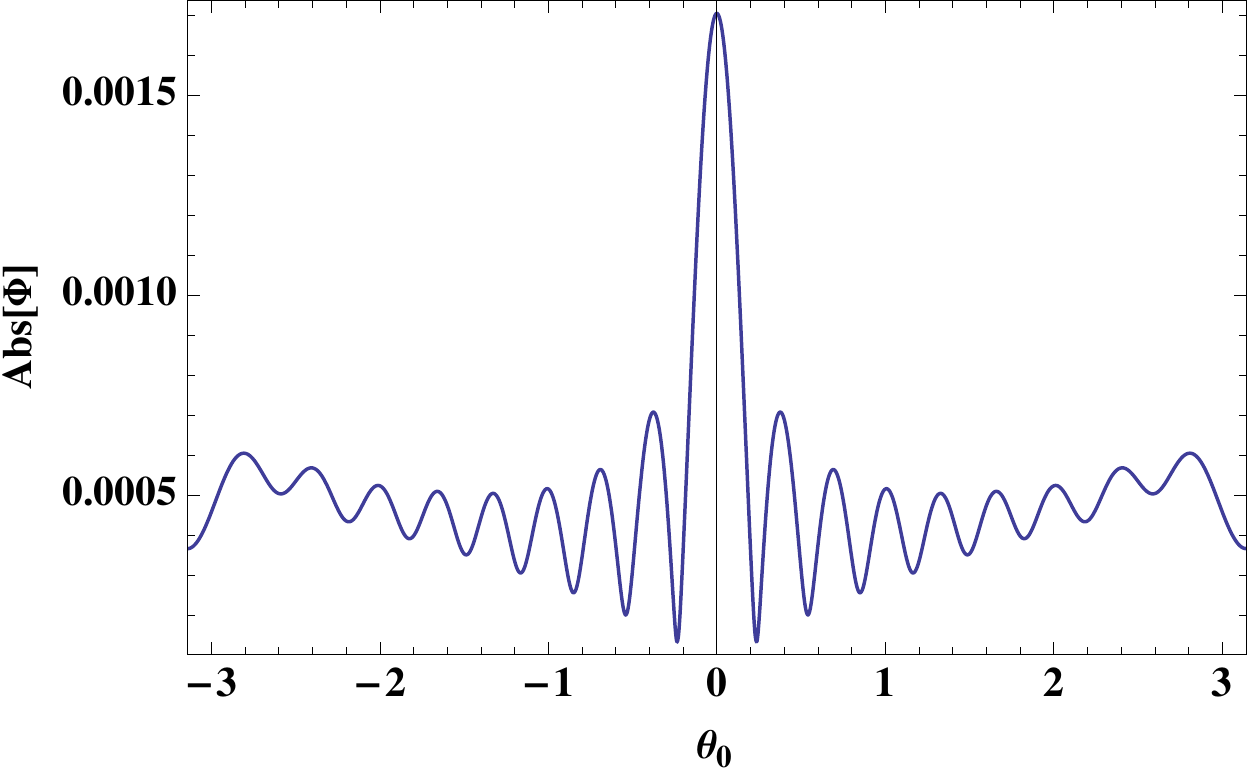}
  \caption{The scattering amplitude at $r_\text{obs}=20M$. $M\omega=2,
  r_\text{S}=2.5M$.}
\end{figure}

\begin{figure}[H]
  \centering
  \includegraphics[width=0.3\linewidth,clip]{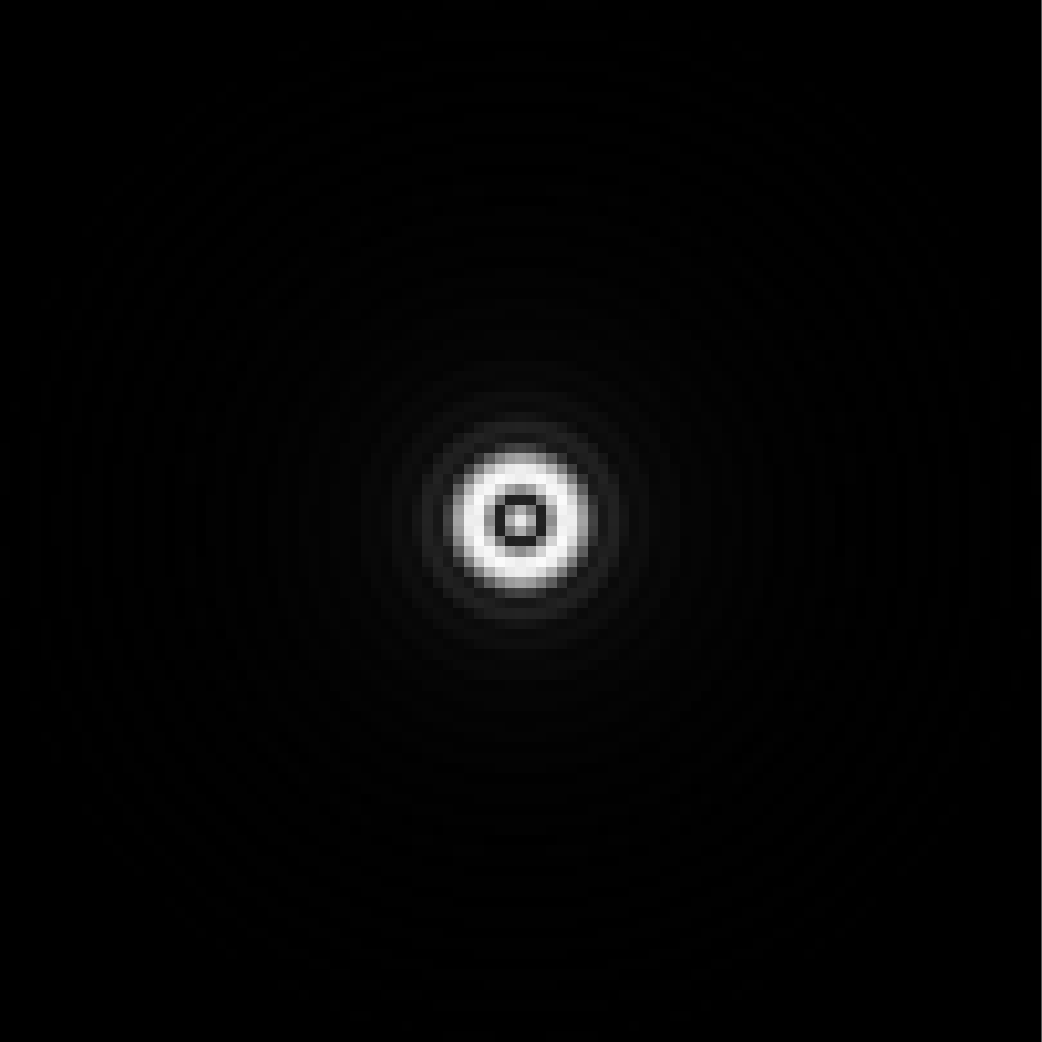}
  \includegraphics[width=0.3\linewidth,clip]{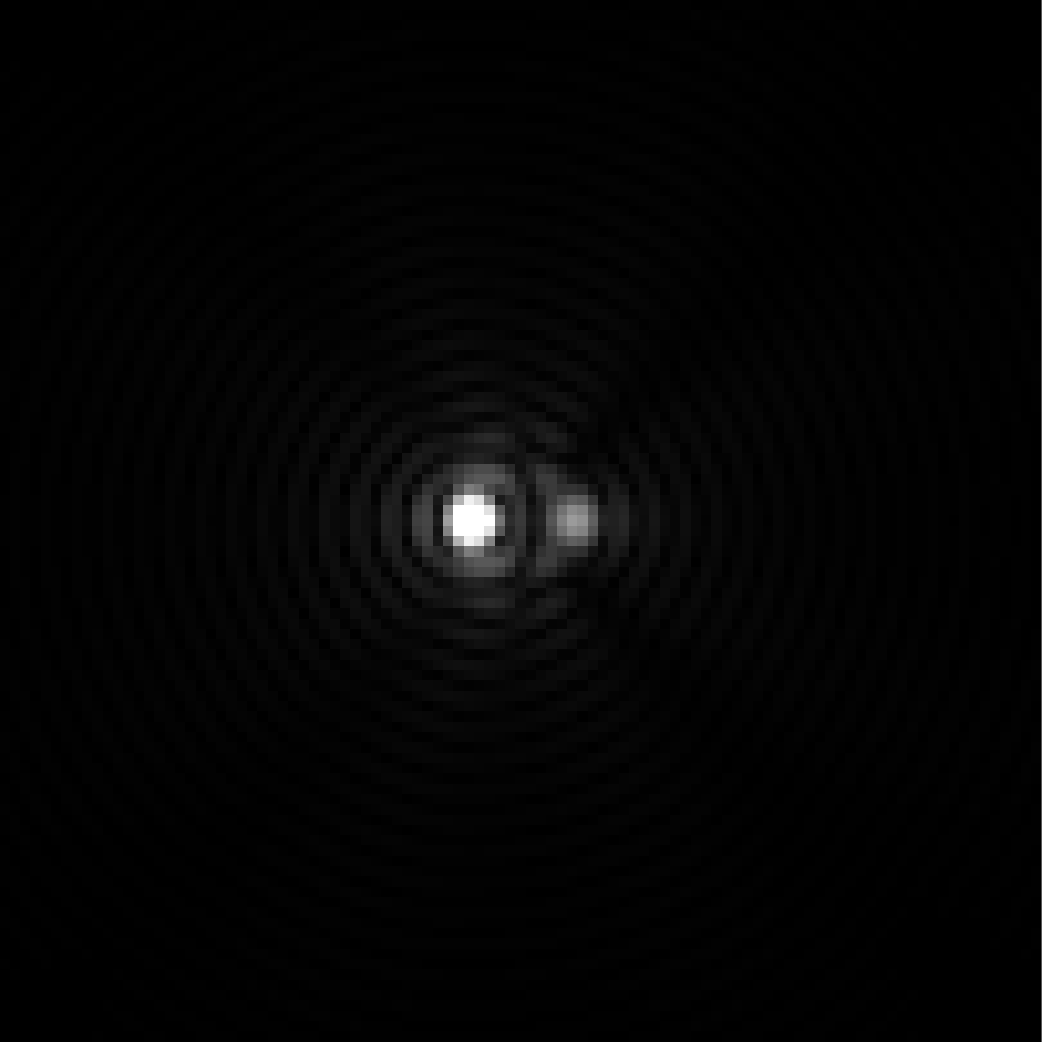}
  \includegraphics[width=0.3\linewidth,clip]{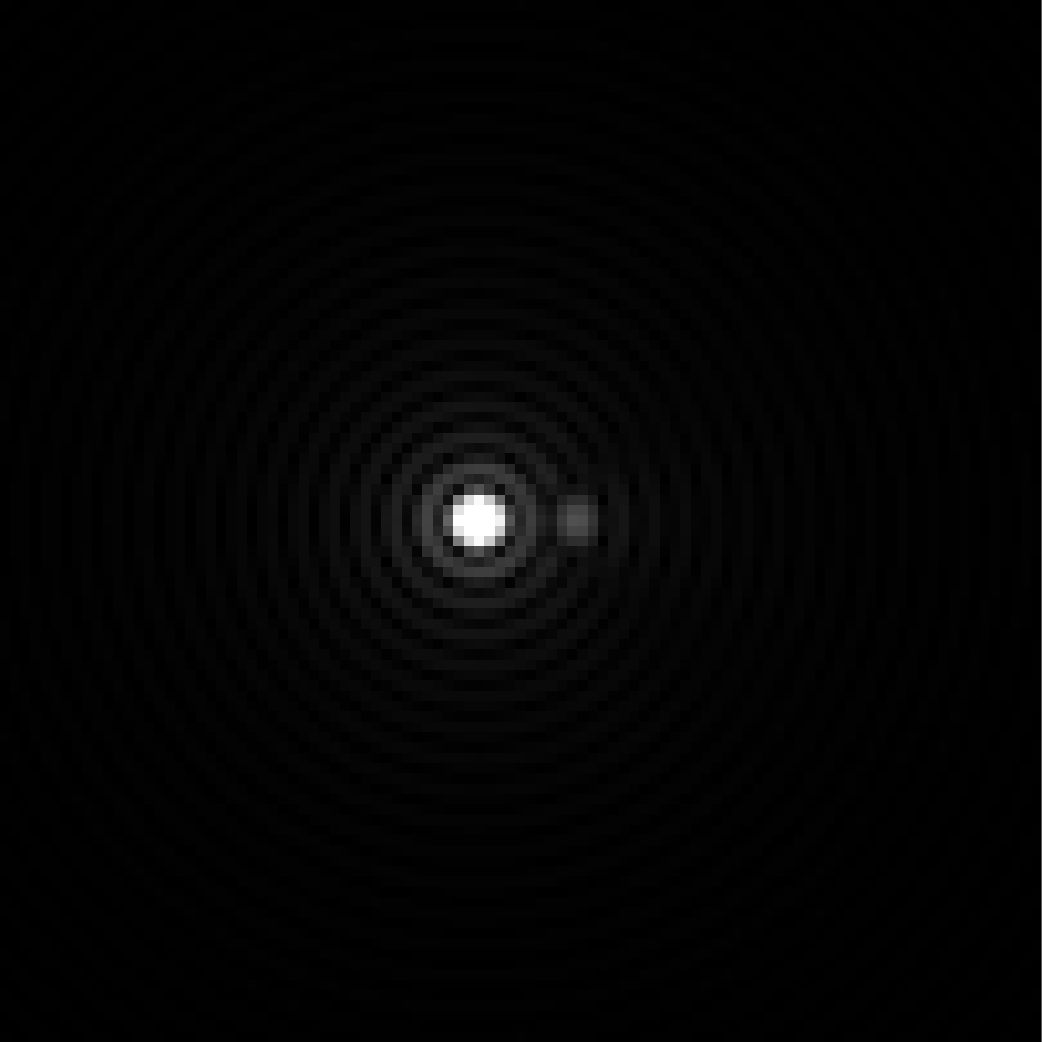}
  \includegraphics[width=0.3\linewidth,clip]{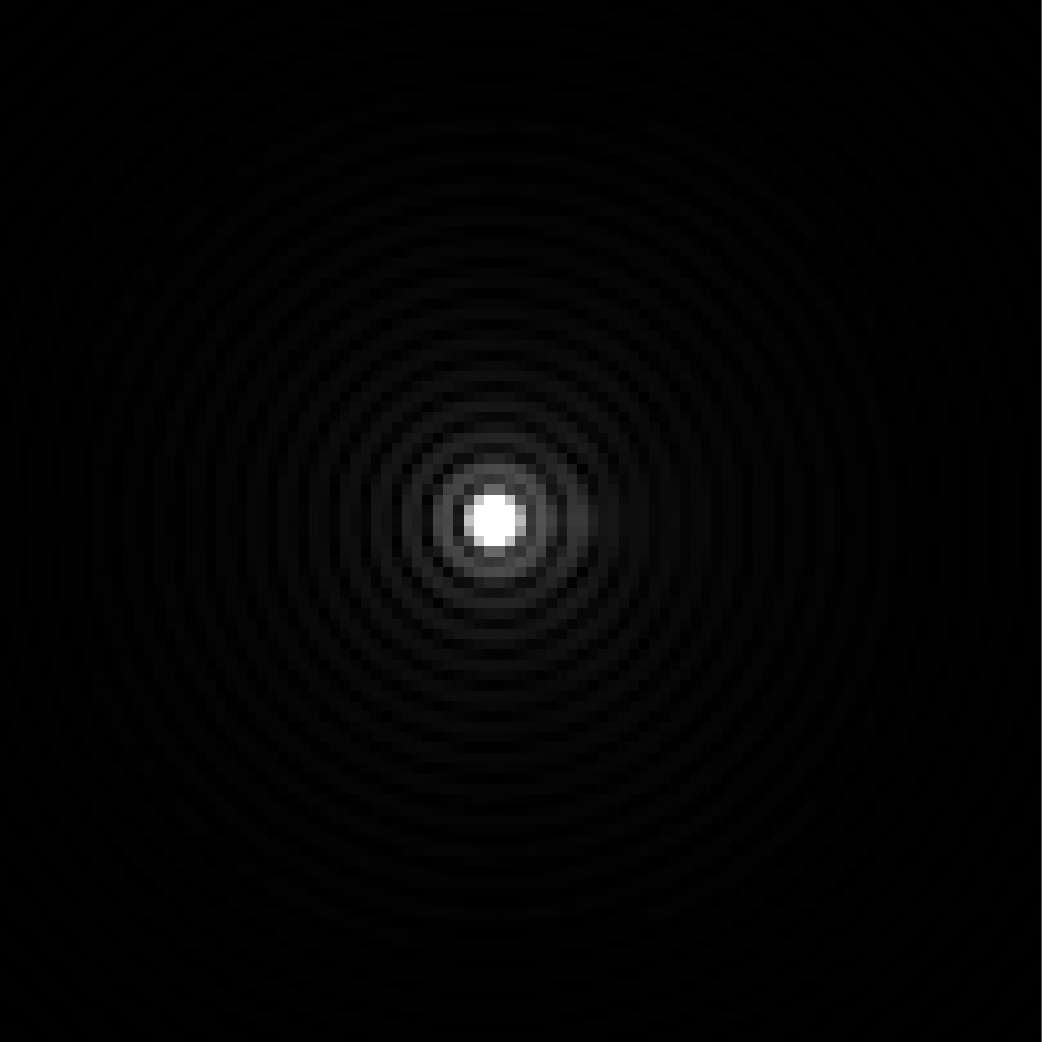}
  \includegraphics[width=0.3\linewidth,clip]{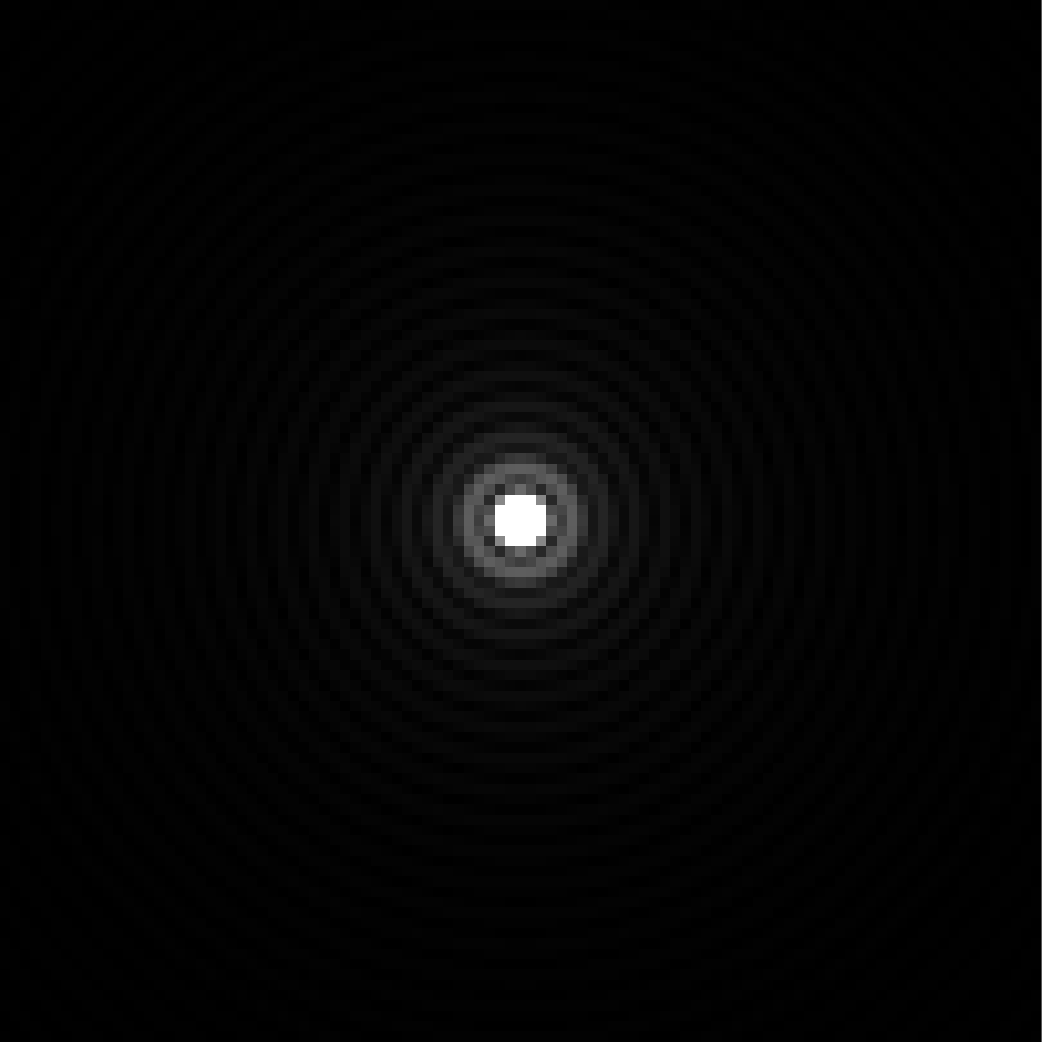}
  \caption{Images of black holes reconstructed from scattering
    waves. From the top left to the bottom right panel, the scattering
    angles are $\theta_0=0,\pi/4,\pi/2,3\pi/4,\pi$.}
\end{figure}

\begin{figure}[H]
  \centering
  \includegraphics[width=0.4\linewidth,clip]{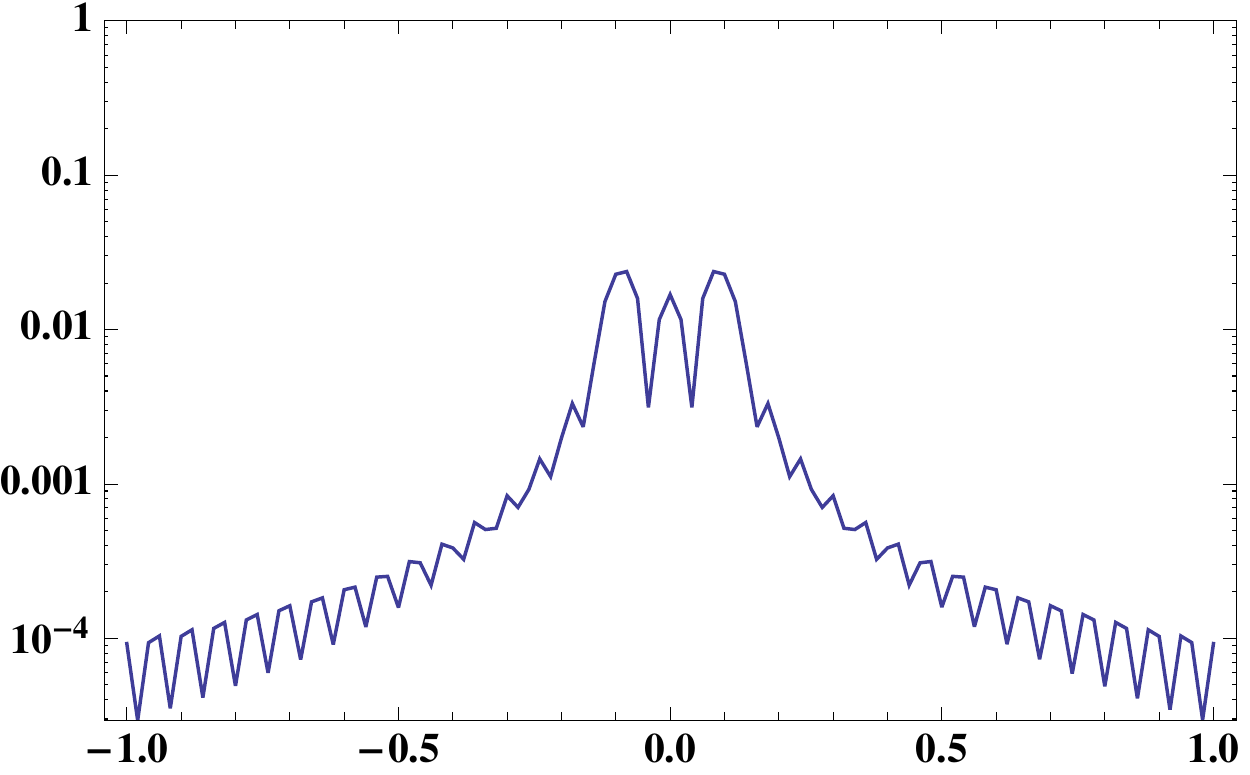}
  \includegraphics[width=0.4\linewidth,clip]{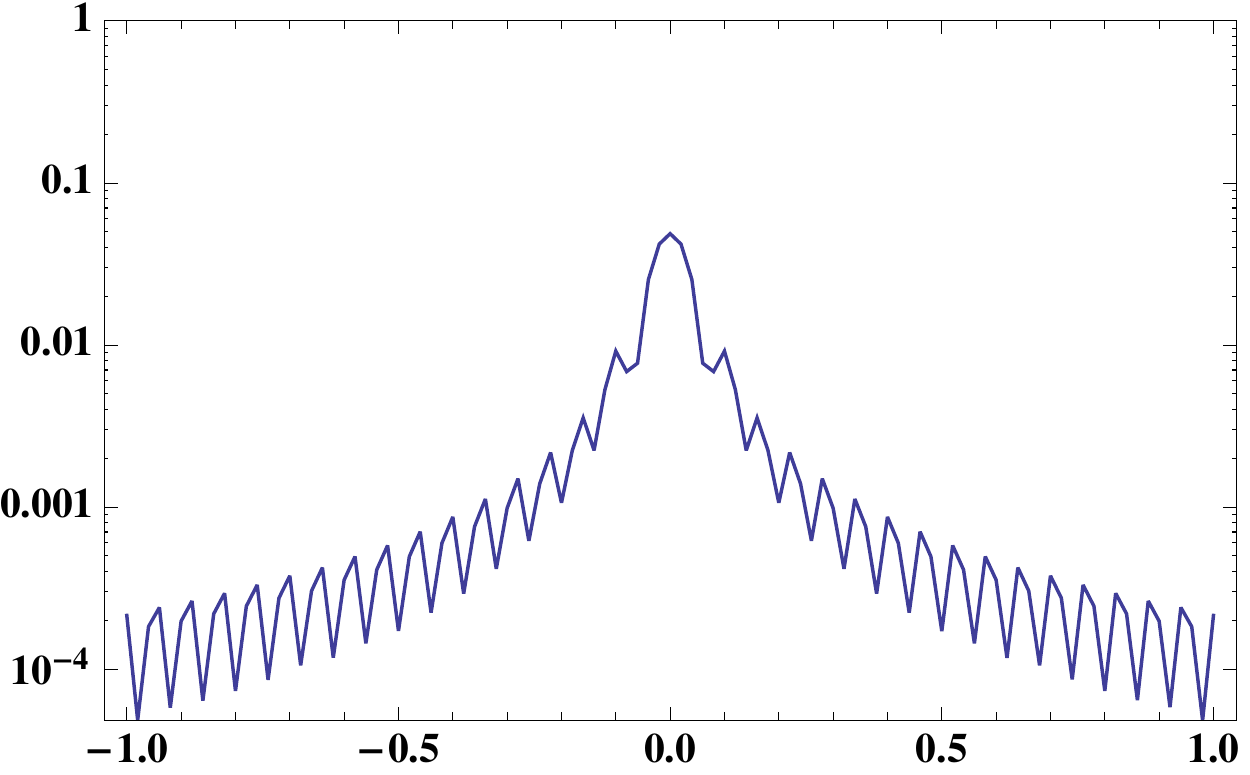}
  \caption{The intensity distribution of images. Left panel:
    $\theta_0=0$. Right panel: $\theta_0=\pi$.}
\end{figure}

\section{Summary and discussion}

We investigated scattering of scalar waves by the Schwarzschild black
hole. Our main aim of this analysis is to obtain images of black
holes using waves. For this purpose, we solved the scalar wave
equation in the Schwarzschild spacetime numerically and reconstructecd
images of black holes by the Fourier transformation of the scattering
waves. For the forward and the backward direction of the scattering,
we obtained ring images corresponding to the glory scattering by the
black hole. In the geometric optics limit, these ring images are
related to existence of the unstable orbit for null rays.  As
extension of analysis presented in this paper, wave scattering and
image formation in the Kerr geometry is an interesting subject to be
tackled. For the Kerr black hole, due to dragging of the spacetime,
incident waves can be amplified by the
superradiance~\cite{FrolovYP:1998}. By investigating images from
scattered waves by the Kerr black hole, we expect to find out a new
aspect and an interpretation of the superradiance in connection with
the Penrose process. We will report on the analysis of the Kerr black
hole case in our next publication.

Another application of the analysis presented in this paper is related
to observations of black hole
shadows~\cite{FalckeH:AJ528:2000,MiyoshiM:PTPS155:2004}. As the
apparent angular sizes of black hole shadows are so small, the
diffraction effect on images are crucial to resolve black hole shadows
in observation using radio interferometer. For SgrA*, which is the
black hole candidate at Galactic center, the apparent angular size of
its shadow is estimated to be $\sim 30\mu$ arcsec and this value
is the largest among black hole candidates.  The resolving power of
the image formation system is given by~\cite{SharmaKK:AP:2006}
\begin{equation}
  \label{eq:power}
  \theta_0=\frac{\lambda}{D},  
\end{equation}
where $\lambda$ is the wave length and $D$ is the size of
``lens''. For the radio interferometer, $D$ corresponds to the
baseline length between antennas. For a sub-mm VLBI, $\lambda\sim
0.1\text{mm}$ and the condition $\theta_0<30 \mu~\text{arcsec}$ yields
$D>1000\text{km}$. This requirement for the baseline length shows the
possibility of detecting the black hole shadow of SgrA$^*$ using the
present day technology of VLBI telescope.  Thus, analysis of black
hole shadows based on wave optics is an important theme for sucessful
detection of shadows and detemination of black hole parameters via
imaging of black holes.

\begin{acknowledgments}
This work was supported in part by the JSPS Grant-In-Aid for
Scientific Research (C) (23540297). The authors thank all member of
``black hole horizon project meeting'' in which the preliminary version of
this paper was presented.
\end{acknowledgments}

\appendix*
\section{finite difference method}

To solve the Helmholtz equation \eqref{eq:rt-wavefunc} numerically, we
should rewrite the equation into the corresponding difference
equation.  For our numerical calculation, we divide the considering
spatial region in $(x,\theta)$ space into $(N+1)\times (N+1)$
homogeneous grids.  The coordinate variables are
\begin{equation}
x_{i}=x_{0}+i\cdot\Delta,\quad
\theta_j=j\cdot\Delta_\theta,\quad i,j=0,\cdots,N,
\end{equation}
where $\Delta$ and $\Delta_\theta$ are the spacing of the grids
in $x$ and $\theta$ coordinates, respectively. The coordinates range
of the numerical box is
\begin{equation}
    x_0\le x\le x_0+N\Delta,\quad 0\le\theta\le N\Delta_\theta\equiv\pi.
\end{equation}
Accordingly, the field $\hat\Phi$ at
the point $(x_{i},\theta_j)$ is written as
$\hat\Phi_{i,j}\equiv\hat\Phi(x_{i},\theta_j)$. Then,
Eq.~(\ref{eq:rt-wavefunc}) in the finite difference form is given by
\begin{align}
&\frac{\hat\Phi_{i+1,j} +
  \hat\Phi_{i-1,j}-2\hat\Phi_{i,j}}{\Delta^2}+\frac{1}{
  r_i^2}\left(1-\frac{2}{
      r_{i}}\right)\left(\cot\theta_j\frac{\hat\Phi_{i,j+1}
      -\hat\Phi_{i,j-1}}{2\delta\Delta}
    +\frac{\hat\Phi_{i,j+1}+\hat\Phi_{i,j-1}
      -2\hat\Phi_{i,j}}{\delta^2\Delta^2}\right)\nonumber\\
&\qquad\qquad\qquad\qquad
+\left[M^2\omega^2-\frac{2}{r_i^3}
\left(1-\frac{2}{r_{i}}\right)\right]\hat\Phi_{i,j}=S_{i,j}
\label{eq:mi-wavefunc}
\end{align}
and Eq.~(\ref{eq:rt-wavefunc2}) is rewritten as
\begin{align}
&\frac{\hat\Phi_{i+1,j}+\hat\Phi_{i-1,j}
  -2\hat\Phi_{i,j}}{\Delta^2}
+\frac{2}{r_i^2}\left(1-\frac{2}{r_i}\right)
\frac{\hat\Phi_{i,j+1}+\hat\Phi_{i,j-1}-2\hat\Phi_{i,j}}{\delta^2\Delta^2}
\nonumber \\
&\qquad\qquad\qquad\qquad
+\left[M^2\omega^2-\frac{2}{r_i^3}\left(1-\frac{2}{r_i}\right)\right]
\hat\Phi_{i,j}=S_{i,j}
\label{eq:mi-wavefunc3}
\end{align}
where we have introduced $\delta=\Delta_\theta/\Delta$ and $r_i$
is defined as $r_i=r(x_{i})$. In these equations, the radial
coordinate is measured in the unit of the black hole mass $M$.

The boundary conditions for the $r$ direction of the Helmholtz
equation must be imposed using Eq.~(\ref{eq:horizon-boundary}) and
Eq.~(\ref{eq:out-boundary}) to determine $\hat\Phi_{-1,j}$ and
$\hat\Phi_{N+1,j}$. These equations are conditions for temporal
behavior of waves. Thus we have to translate these conditions for the
wave field to that does not contain the time dependence.  We first
consider Eq.~(\ref{eq:horizon-boundary}). This equation relates the
wave field at the inner spatial boundary $x_0$ and $x_{-1}$. As the
solution of this equation is $\hat\Phi(t,x)=\hat\Phi(t+x)$,
the field value $\hat\Phi(t+\Delta_t,x_{-1})$ is expressed as
\begin{equation}
\hat\Phi(t+\Delta_t,x_{-1})=\hat\Phi(t,x_{-1}+\Delta_t).
\end{equation}
As the time dependence of the wave is assumed to be
$\hat\Phi(t)\propto e^{-iM\omega t}$, this boundary condition
\eqref{eq:horizon-boundary} provides the following relation between the
field values at $x_0$ and $x_{-1}$:
\begin{equation}
e^{iM\omega \Delta_t}\hat\Phi_{-1,j}
=\hat\Phi_{-1,j}+\frac{\Delta_t}{\Delta}(\hat\Phi_{0,j}-\hat\Phi_{-1,j}).
\end{equation}
Thus, $\hat\Phi_{-1,j}$ can be determined by
\begin{equation}
\hat\Phi_{-1,j}
=\frac{-\delta_t}{1-\delta_t-e^{-iM\omega\delta_t\Delta}}\hat\Phi_{0,j}
\label{eq:mi-boundary1}
\end{equation}
where $\delta_t\equiv\Delta_t/\Delta$.  The boundary condition
at the outer spatial boundary can be obtained in the similar way and we
have 
\begin{equation}
\hat\Phi_{N+1,j}
=\frac{-\delta_t}{1-\delta_t-e^{-iM\omega\delta_t\Delta}}\hat\Phi_{N,j}.
\label{eq:mi-boundary2}
\end{equation}
In our numerical analysis, we set $\delta_t=1$ and we adopt the
following equations to impose the boundary conditions at the inner and the
outer spatial boundaries:
\begin{equation}
\hat\Phi_{-1,j}=e^{iM\omega\Delta}\hat\Phi_{0,j},\quad
\hat\Phi_{N+1,j}=e^{iM\omega\Delta}\hat\Phi_{N,j}.
\label{eq:boundary-condition1}
\end{equation}
The boundary conditions on the $\bar{z}$ axis are determined by
Eq.~(\ref{eq:theta-boundary}) and
\begin{equation}
\hat\Phi_{i,-1}=\hat\Phi_{i,1},\quad
\hat\Phi_{i,N+1}=\hat\Phi_{i,N-1}.
\label{eq:mi-boundary4}
\end{equation}

In the numerical grid space, the location of
the point source is $(x_{i_\text{S}},\pi)$. Equation \eqref{eq:waveS}
yields field values at
points $(x_{i_\text{S}-1},\pi), (x_{i_\text{S}+1},\pi)$ and
$(x_{i_\text{S}},\pi-\Delta_\theta)$:
\begin{equation}
  \label{eq:wsource}
  \hat\Phi_{i_\text{S}-1,N}=\hat\Phi_{i_\text{S}+1,N}
  =\frac{A\,r_\text{S}e^{i\omega\Delta}}{\Delta\sqrt{1-2M/r_\text{S}}},\quad
  \hat\Phi_{i_\text{S},N-1}=\frac{A}{\Delta_\theta}e^{i\omega
    r_\text{S}\Delta_\theta/\sqrt{1-2M/r_\text{S}}}.
\end{equation}
\eqref{eq:wsource}  provides the boundary condition for the wave equation
around the point source.

These difference equations and the boundary conditions constitute the
simultaneous equations. The simultaneous equations are written in a
matrix form as
\begin{equation}
\bs{A}\cdot\bs{x}=\bs{b}
\label{eq:matrix1}
\end{equation}
where $\bs{x}$ is the vector to be solved and the matrix $\bs{A}$
and the vector $\bs{b}$ are determined by the difference equation.
As  demonstrations, let us first consider the $3\times 3$ grids case.
The components of $\bs{A}$, $\bs{x}$ and $\bs{b}$ are 
\begin{equation}
\bs{A}=
\begin{pmatrix}
\bs{C}_1 & \bs{D}_1 & \bs{0} \\
\bs{H}_1 & \bs{C}_2 & \bs{H}_1 \\
\bs{0} & \bs{D}_1 & \bs{C}_1 \\
\end{pmatrix} 
,\quad
\bs{x}=
\begin{pmatrix}
\hat\Phi_{0,0}^{R}\\
\hat\Phi_{0,0}^I\\
\hat\Phi_{1,0}^R\\
\hat\Phi_{1,0}^I\\
\vdots\\
\hat\Phi_{2,2}^R\\
\hat\Phi_{2,2}^I\\
\end{pmatrix}
,
\quad
\bs{b}=
\begin{pmatrix}
S^R_{0,0}\\
S^I_{0,0}\\
S^R_{1,0}\\
S^I_{1,0}\\
\vdots\\
S^R_{2,2}\\
S^I_{2,2}\\
\end{pmatrix},
\quad
\end{equation}
where $\hat\Phi^R$ and $\hat\Phi^I$ represent the real and the
imaginary part of $\hat\Phi$, respectively. $\bs{C}_1, \bs{C}_2,
\bs{D}_1$ and $\bs{H}_1$ are $6\times 6$ matrices given by
\begin{equation}
\bs{C}_1=
\begin{pmatrix}
c_0 & -a_0 & r_0 ^4\delta^2 & 0 & 0 & 0 \\
a_0 & c_0 & 0 & r_0 ^4\delta^2 &0 & 0 \\
r_1^4\delta^2 & 0 & d_1 & 0 & r_1^4\delta^2 & 0 \\
0 & r_1^4\delta^2 & 0 & d_1 & 0 & r_1^4\delta^2 \\
0 & 0 & r_2^4\delta^2 & 0 & c_2 & -a_2\\
0 & 0 & 0 & r_2^4\delta^2 & a_2 & c_2\\
\end{pmatrix} ,
\end{equation}
\begin{align*}
a_i&=r_i^4\delta^2\sin(M\omega\Delta),\nonumber \\
c_i&=\delta^2\Delta^2\left(M^2\omega^2r_i^4-2(r_i-2)\right)-2r_i^4\delta^2
-4r_i(r_i-2)+r_i\delta^2\cos(M\omega\Delta),\nonumber \\
d_i&=\delta^2\Delta^2\left(M^2\omega^2r_i^4-2(r_i-2)\right)-2r_i^4\delta^2
-4r_i(r_i-2),\nonumber
\end{align*}
\begin{equation}
\bs{C}_2=
\begin{pmatrix}
h_0 & -2a_0 & 2r_0 ^4\delta^2 & 0 & 0 & 0 \\
2a_0 & h_0 & 0 & 2r_0 ^4\delta^2 &0 & 0 \\
2r_1^4\delta^2 & 0 & k_1 & 0 & 2r_1^4\delta^2 & 0 \\
0 & 2r_1^4\delta^2 & 0 & k_1 & 0 & 2r_1^4\delta^2 \\
0 & 0 & 2r_2^4\delta^2 & 0 & h_2 & -2a_2\\
0 & 0 & 0 & 2r_2^4\delta^2 & 2a_2 & h_2\\
\end{pmatrix},
\end{equation}
\begin{align*}
h_i&=2\delta^2\Delta^2\left(M^2\omega^2r_i^4-2(r_i-2)\right)
-4r_i^4\delta^2-4r_i(r_i-2)+2r_i\delta^2\cos(M\omega\Delta),\nonumber \\
k_i&=2\delta^2\Delta^2\left(M^2\omega^2r_i^4-2(r_i-2)\right)
-4r_i^4\delta^2-4r_i(r_i-2),\nonumber
\end{align*}
\begin{align}
&\bs{D}_1=\mathrm{diag}\Bigl[4r_0(r_0-2),4r_0(r_0-2),4r_1(r_1-2),4r_1(r_1-2),
4r_2(r_2-2),4r_2(r_2-2)\Bigr], \\
&\bs{H}_j=\mathrm{diag}\Bigl[r_0(r_0-2)(2-\delta\Delta\cot\theta_j),
r_0(r_0-2)(2-\delta\Delta\cot\theta_j),r_1(r_1-2)(2-\delta\Delta\cot\theta_j),
\nonumber \\
&\qquad\qquad
r_1(r_1-2)(2-\delta\Delta\cot\theta_j),r_2(
r_2-2)(2-\delta\Delta\cot\theta_j),r_2(
r_2-2)(2-\delta\Delta\cot\theta_j)
\Bigr]. 
\end{align}
In the case of $4\times 4$ grids, the structure of $\bs{x}$ and $\bs{A}$
becomes as follows:
\begin{equation}
\bs{x}=
\begin{pmatrix}
\hat\Phi_{0,0}^{R}\\
\hat\Phi_{0,0}^I\\
\hat\Phi_{1,0}^R\\
\hat\Phi_{1,0}^I\\
\vdots\\
\hat\Phi_{3,3}^R\\
\hat\Phi_{3,3}^I\\
\end{pmatrix}
,\quad
\bs{A}=
\begin{pmatrix}
\bs{C}_1 & \bs{D}_1 & \bs{0} & \bs{0} \\
\bs{H}_1 & \bs{C}_2 & \bs{H}_1 & \bs{0}\\
\bs{0} & \bs{H}_2 & \bs{C}_2 & \bs{H}_2 \\
\bs{0} & \bs{0} & \bs{D}_1 & \bs{C}_1
\end{pmatrix} .
\label{eq:A}
\end{equation}
$\bs{C}_{1,2}, \bs{D}_1$ and $\bs{H}_{1,2}$ are $8\times 8$ matrices
given by
\begin{equation}
\bs{C}_1=
\begin{pmatrix}
c_0 & -a_0 & r_0 ^4\delta^2 & 0 & 0 & 0 & 0 & 0\\
a_0 & c_0 & 0 & r_0 ^4\delta^2 &0 & 0 & 0 & 0\\
r_1^4\delta^2 & 0 & d_1 & 0 & r_1^4\delta^2 & 0 & 0 & 0\\
0 & r_1^4\delta^2 & 0 & d_1 & 0 & r_1^4\delta^2 & 0 & 0\\
0 & 0 & r_2^4\delta^2 & 0 & d_2 & 0 & r_2^4\delta^2 & 0\\
0 & 0 & 0 & r_2^4\delta^2 & 0 & d_2 & 0 & r_2^4\delta^2\\
0 & 0 & 0 & 0 & r_3^4\delta^2 & 0 & c_3 & -a_3\\
0 & 0 & 0 & 0 & 0 & r_3^4\delta^2 & a_3 & c_3\\
\end{pmatrix},
\end{equation}
\begin{equation}
\bs{C}_2=
\begin{pmatrix}
h_0 & -2a_0 & 2r_0 ^4\delta^2 & 0 & 0 & 0 & 0 & 0\\
2a_0 & h_0 & 0 & 2r_0 ^4\delta^2 &0 & 0 & 0 & 0\\
2r_1^4\delta^2 & 0 & k_1 & 0 & 2 r_1^4\delta^2 & 0 & 0 & 0\\
0 & 2 r_1^4\delta^2 & 0 & k_1 & 0 & 2r_1^4\delta^2 & 0 & 0\\
0 & 0 & 2r_2^4\delta^2 & 0 & k_2 & 0 & 2r_2^4\delta^2 & 0\\
0 & 0 & 0 & 2r_2^4\delta^2 & 0 & k_2 & 0 & 2r_2^4\delta^2\\
0 & 0 & 0 & 0 & 2r_3^4\delta^2 & 0 & h_3 & -2a_3\\
0 & 0 & 0 & 0 & 0 & 2r_3^4\delta^2 & 2a_3 & h_3\\
\end{pmatrix},
\end{equation}
\begin{align}
&\bs{D}_1=\mathrm{diag}\Bigl[
4r_0(r_0-2),4r_0(r_0-2),4r_1(r_1-2),4r_1(r_1-2),4r_2(r_2-2)
,4r_2(r_2-2) \\
&\qquad\qquad\qquad\qquad
,4r_3(r_3-2),4r_3(r_3-2)\Bigr], \notag\\
&\bs{H}_j=\mathrm{diag}\Bigl[
r_0(r_0-2)(2-\delta\Delta\cot\theta_j),r_0(r_0-2)(2-\delta\Delta\cot\theta_j),
r_1(r_1-2)(2-\delta\Delta\cot\theta_j), \nonumber\\
&\qquad\qquad r_1(r_1-2)(2-\delta\Delta\cot\theta_j),
r_2(r_2-2)(2-\delta\Delta\cot\theta_j),
r_2(r_2-2)(2-\delta\Delta\cot\theta_j),\\
&\qquad\qquad r_3(r_3-2)(2-\delta\Delta\cot\theta_j),
r_3(r_3-2)(2-\delta\Delta\cot\theta_j)].\nonumber
\end{align}
If the number of the grids becomes larger, it is possible to predict
the structure of $\bs{A}$ from these examples. Of course, we should modify the
forms of $\bs{A}$ and $\bs{b}$ to introduce a source of waves along the
method we have described in \eqref{eq:wsource}.

In our calculation, the region of the numerical box is $-2\le x\le 23
~(2.03\leq r\leq20.5),~0\leq\theta\leq\pi$ and we divide this region
into $1001\times1001$ grids.  We used the Mathematica to obtain the
solution of the linear system \eqref{eq:matrix1}.

\end{document}